\documentclass[pra,twocolumn,floatfix]{revtex4-2}

\usepackage[utf8]{inputenc}
\usepackage{xcolor}
\usepackage[backref=none]{hyperref}
\hypersetup{
    colorlinks=true,  
    linkcolor=blue,   
    citecolor=blue,   
    urlcolor=olive,   
}

\usepackage{amsmath}
\usepackage{amssymb}
\usepackage{multirow}
\usepackage{graphicx}
\graphicspath{ {images/} }
\usepackage{array}
\usepackage{upgreek}
\usepackage{mathtools}
\usepackage{cuted}

\usepackage[switch]{lineno}

\makeatletter

\begin{document}

\title{Loosely Bound Few-Body States in a Spin-1 Gas with Near-Degenerate Continua}

\author{Yaakov Yudkin$^{1}$}
\author{Paul S. Julienne$^{2}$}
\author{Lev Khaykovich$^{1}$}

\affiliation{$^{1}$Department of Physics and QUEST Center and Institute of Nanotechnology and Advanced Materials, Bar-Ilan University, Ramat-Gan 5290002, Israel}

\affiliation{$^{2}$Joint Quantum Institute (JQI), University of Maryland and NIST, College Park, Maryland 20742, USA}

\date{\today}

\begin{abstract}
A distinguishing feature of ultracold collisions of bosonic lithium atoms is the presence of two near-degenerate two-body continua.
The influence of such a near-degeneracy on the few-body physics in the vicinity of a narrow Feshbach resonance is investigated within the framework of a minimal model with two atomic continua and one closed molecular channel.
The model allows analysis of the spin composition of loosely bound dimers and trimers.
In the two-body sector the well-established coupled-channels calculations phenomenology of lithium is qualitatively reproduced, and its particularities are emphasized and clarified.
In the three-body sector we find that the Efimov trimer energy levels follow a different functional form as compared to a single continuum scenario while the thresholds remain untouched.
This three-channel model with two atomic continua complements our earlier developed three-channel model with two molecular channels~\cite{Yudkin21} and suggests that the experimentally observed exotic behavior of the first excited Efimov energy level~\cite{Yudkin20} is most probably caused by the short-range details of the interaction potential.
\end{abstract}

\maketitle

\section{Introduction}

Two decades of experiments with ultracold atoms have stimulated spectacular advance in our understanding of few-body systems.
Especially intriguing is scattering in the vicinity of a Feshbach resonance~\cite{Chin10}, near a so-called singularity, because diverging two-body interactions give rise to a large variety of bound clusters whose properties are universal functions of the scattering length~\cite{Braaten&Hammer06,Naidon17,Greene17,D'Incao18}.
Several experimental and theoretical works were able to establish this universality~\cite{Zaccanti09,Ferlaino09,Huang14,Xie20}.

Apart from the scattering length $a$, the systems' size and energy length scales are governed by the three-body parameter which, in the case of short-range two-body interaction potentials, succumbs to another type of universality.
This Efimov-van der Waals universality relates the three-body parameter to the van der Waals length $R_{\rm vdW}$ of the underlying two-body potential~\cite{Berninger11,Wang12,Naidon14}.
While this holds for broad resonances~\cite{Wild12,Tung14,Zenesini14,Fletcher17}, in the vicinity of narrow resonances the three-body parameter was shown to deviate from such universality~\cite{Ji15,Johansen17,Chapurin19,Li22,Etrych22}.

\begin{figure}
\includegraphics[width=\linewidth]{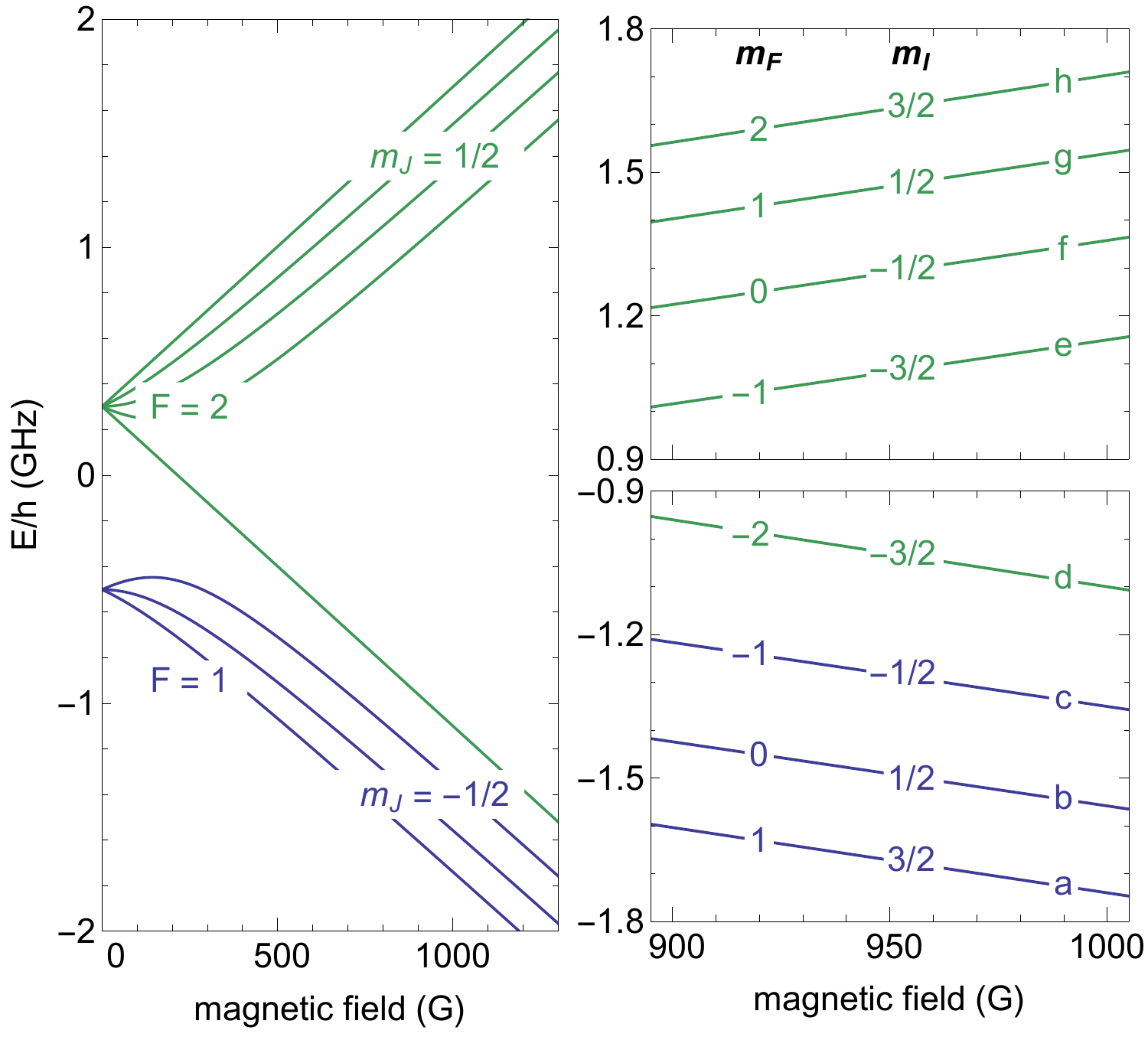}
\caption{\label{fig:Li7:Zeeman shift}
Plot of the energy of the hyperfine levels of the $^7$Li 2S$_{1/2}$ ground state versus magnetic field $B$ in Gauss.
The labels of the levels used throughout this paper are shown.}
\end{figure}

Although the concept of universality is now well understood, bosonic lithium ($^7$Li) deviates from it in various aspects and continues to puzzle experimentalists and theorists.
For one, $^7$Li satisfies the Efimov-van der Waals universality quite well even though all its resonances are narrow~\cite{Gross11}.
Further, it was shown that modelling experimental results requires relatively complex multi-channel theories~\cite{Wang14} with, on occasion, even energetically distant channels playing a crucial role~\cite{Secker21}.
In addition, it has been speculated that generic three-body forces must be taken into account to achieve quantitative agreement between experiment and theory~\cite{D'IncaoPrivate}.
Finally, the spectrum of the timer was observed to cross into the atom-dimer continuum (or go through an avoided crossing) instead of merging with it~\cite{Yudkin20}.

While the typical few-body experiment is conducted with atoms polarized in the absolute ground state (the $a$-state), in $^7$Li the second lowest state ($b$-state, see Fig.~\ref{fig:Li7:Zeeman shift}) was shown to be extraordinarily stable against dipolar relaxation~\cite{Gross11}.
Bosonic lithium is thus an attractive species for experiments with either spin state or a mixture of both~\cite{Jepsen22} .
In particular, it provides an opportunity to compare the few-body physics, e.g. the three-body parameter and the Efimov spectrum, of the two channels~\cite{Gross10,Gross11}.

In the case of $^7$Li (and many other species), Feshbach resonances occur at high magnetic fields $B$, such that $\mu_BB\gg A_{hf}$ (where $\mu_B$ is the Bohr magneton and $A_{hf}$ the hyperfine-structure constant) and the Zeeman shift $\sim\mu_BBm_J$ dominates over the hyperfine splitting $\sim A_{hf}m_Jm_I$.
Here, $m_J$ and $m_I$ are the electronic angular momentum (spin + orbital) and nuclear spin projections, respectively, and $m_F=m_J+m_I$ (Fig.~\ref{fig:Li7:Zeeman shift}).
Therefore, when considering differences within the $m_J=-1/2$ subset the Zeeman shifts cancel leaving the hyperfine splitting ($\sim A_{hf}/2$) to dominate.
To first order in hyperfine interaction, the differences between the lowest pair ($a$, $b$) and the second lowest pair ($b$, $c$) are identical such that the two-body $bb$ and $ac$ channels are degenerate.
To higher order the degeneracy is lifted and amounts to a few to a few tens of MHz.
Moreover, the total spin projection of the $bb$- and $ac$-channels are identical which permits energy-conserving spin-exchange coupling between the two.
This near-degeneracy of a same-spin state in the $bb$-channel, and its absence in the $aa$-channel, motivates a comparison of the spin composition of the loosely bound states.
Indeed, full two-body coupled-channels calculations show that the near-threshold $bb$ eigenstate has a non-negligible $ac$-population and we explore its implications here by building a simplified model for the $bb$-channel~\cite{bb_channel_meaning}.
Another strong motivation for the model originates in an effort to understand exotic behavior of the first excited Efimov energy level in the vicinity of the atom-dimer threshold observed in a recent experiment~\cite{Yudkin20}.
Although the attempt ultimately fails, it indicates that the solution may be hidden in the short range details of the interatomic interaction potential.

The paper is organized as follows. 
First, the two-body sector observables of $^7$Li are presented in Sec.~\ref{sec:2-body Li-7} and the distinctiveness of the $bb$-channel, in comparison to the $aa$-channel, is demonstrated.
In Sec.~\ref{sec:model + hamiltonian}, we introduce a model for the $bb$-channel which captures the physics with three channels~\cite{bb_channel_meaning}.
The two-body scattering and bound state observables of the model are derived in Sec.~\ref{sec:2-body}.
We compare the results to the standard two-channel model and demonstrate qualitative agreement with full coupled-channels calculations of the $aa$- and $bb$-scattering-channels of $^7$Li.
In Sec.~\ref{sec:3-body} we apply the model to the three-body sector and find that the functional form of the trimer's binding energy is altered.
Although the thresholds are not affected by the $ac$-channel, the binding energy is consistently lower than in its absence.
After following-up with a quick discussion on the lithium few-body puzzle and the implications of the developed model in Sec.~\ref{sec:discussion} we conclude the paper.

Note that together with the previously developed model with one open atomic and two closed molecular channels, used for reproducing overlapping Feshbach resonances~\cite{Yudkin21,Li22}, the current treatment complements the description of the complexity of the $bb$-channel in $^7$Li within the well-defined framework of the simplified theory.
Thus, a meaningful comparison with the experimental results is performed in Sec.~\ref{sec:discussion}.

\section{Li-7}
\label{sec:2-body Li-7}

The $bb$-channel features a Feshbach resonance at $B_{\text{res},bb}=893.78(4)$ G and a second, overlapping, much narrower one at $845.31(4)$ G~\cite{Jachymski13,Julienne14}.
The former is the resonance of interest but the effect of the latter cannot be fully ignored.
Since it was considered in depth in Ref.~\cite{Yudkin21} we use the conclusions of that work and focus on the additional closed atomic $ac$-channel here.

\begin{figure}
\includegraphics[width=\linewidth]{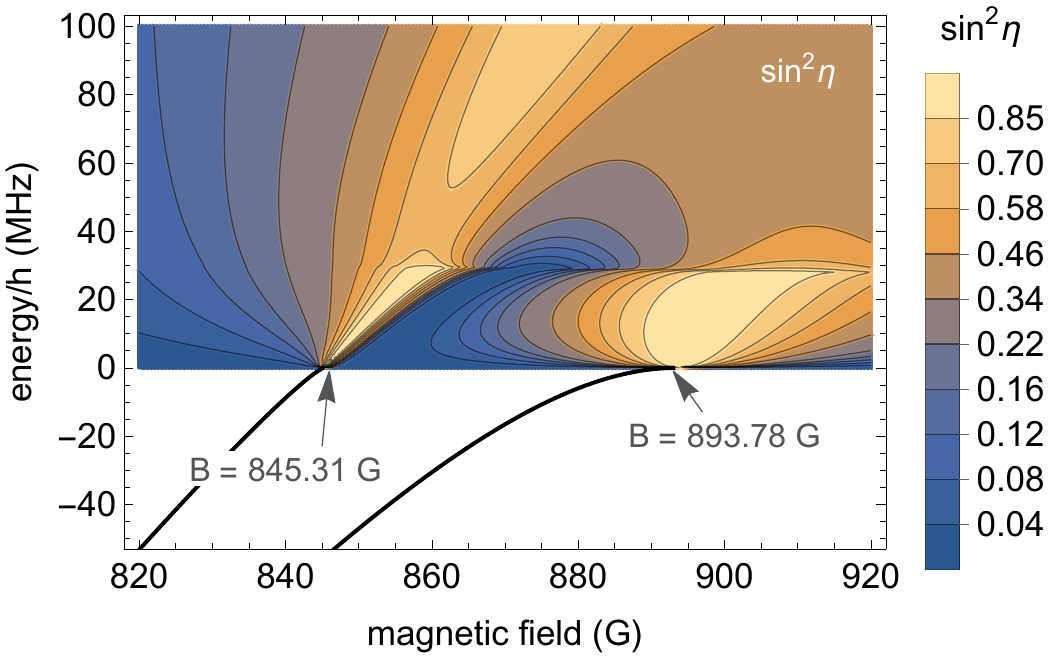}
\caption{\label{fig:Li7:scattering amplitude}
Coupled-channels calculation for two atoms in the $b$-state showing $\sin^2\eta_{bb}(E,B)$ for elastic scattering for $E>0$ and the dimer binding energies for $E<0$.
Here $E=0$ is the energy of two atoms in the $b$ state at magnetic field $B$.
At the $ac$-threshold ($E/h=28$ MHz) the scattering exhibits cusp behavior.
}
\end{figure}

Our coupled-channels calculations are based on the molecular potentials determined in Ref.~\cite{Julienne14} and include all $s$-wave two-body channels with the same total angular momentum projection $M_{\text{tot}}$, which in this case is just the total spin projection summed over both atoms.
Bound or scattering wave functions are written as $|\psi\left(R\right)\rangle=\sum_{ij}\phi_{ij}\left(R\right)|ij\rangle$, where $ij$ denotes the channel indices specifying all atomic states with the same $M_\text{tot}$ included in the coupled-channels basis set, and the components $\phi_{ij}$ reflect closed or open channel boundary conditions depending of the total energy $E$.
Figure~\ref{fig:Li7:scattering amplitude} shows the energies $E<0$ of the two bound states with $M_F=0$ below the $bb$-channel threshold and the elastic scattering indicated by $\sin^2\eta_{bb}(E,B)$ for $E>0$ above the threshold, where $\eta_{bb}(E,B)$ is the asymptotic phase shift for the $bb$-channel.
Strong near-unitary elastic scattering persists above threshold as the two bound states emerge into the $bb$ scattering continuum as resonant features, exhibiting cusp behavior at the $ac$ threshold.

\begin{figure}
\includegraphics[width=\linewidth]{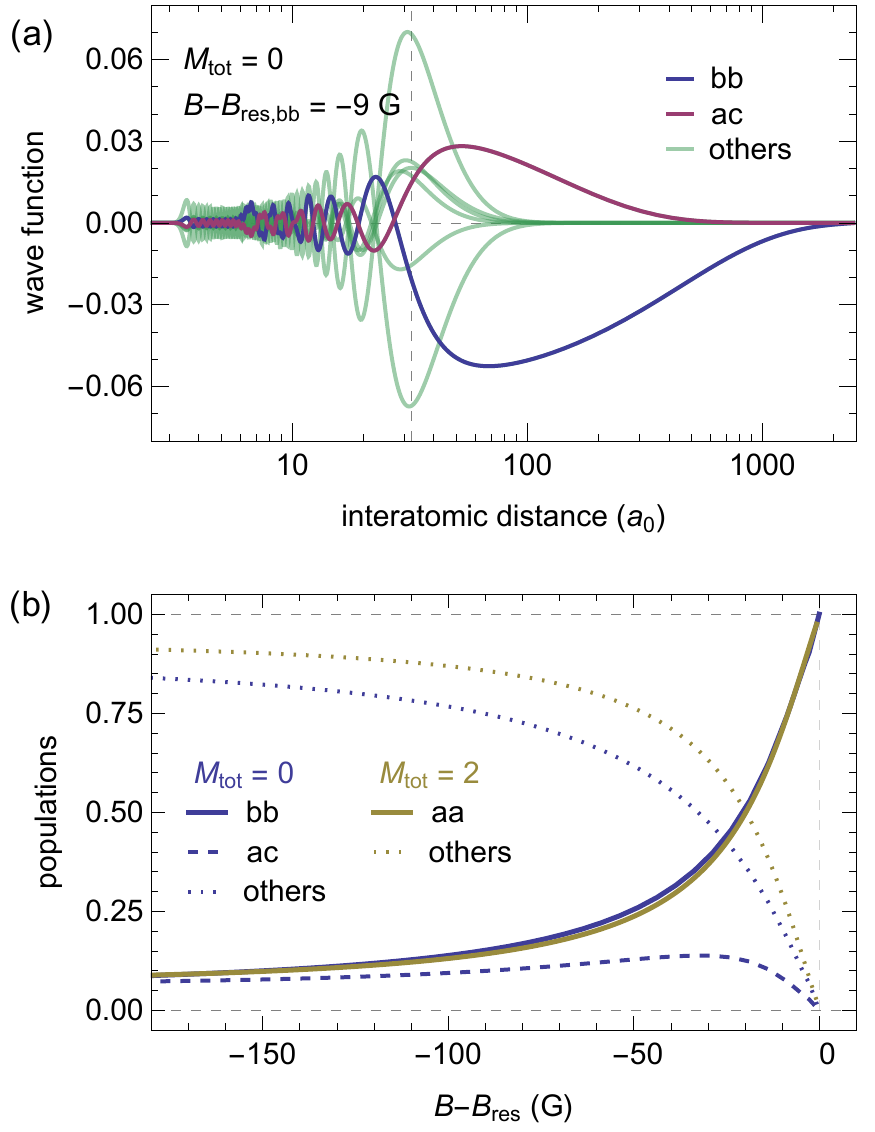}
\caption{\label{fig:Li7:wave functions and population}
(a) Wave function components (see text) of the dimer bound state close to resonance in the basis of the $M_{\text{tot}}=0$ two-body channels, where a$_0 \approx 0.0529$nm is one atomic unit of length.
The vertical dashed line, where all six energetically distant channels (faint green) are maximal, indicates the van der Waals length $R_\text{vdW}=$ 32 a$_0$.
Both the $bb$ and the $ac$ components reach far beyond $R_\text{vdW}$.
Note the logarithmic scale of the horizontal axis.
(b) Magnetic field dependence of the populations.
}
\end{figure}

The eight components of the $M_{\text{tot}}=0$ bound state in the $bb$ channel 9 G below the 893.78G resonance are shown in Fig.~\ref{fig:Li7:wave functions and population}(a).
In addition to the $bb$- and $ac$-channels there are another six two-body channels: $ae$, $bf$, $cg$, $dh$, $eg$ and $ff$ (see Fig.~\ref{fig:Li7:Zeeman shift}).
Around the resonance position ($893.78$ G), the $ac$-channel lies only $\left(E_{ac}-E_{bb}\right)/h=28$ MHz above the $bb$-channel threshold (see Fig.~\ref{fig:Li7:scattering amplitude}), whereas the other six atomic channels include at least one $m_J=+1/2$ spin component such that the Zeeman shift makes them energetically very distant from the $bb$ or $ac$ channels ($\gtrsim2$~GHz for $B>800$~G).
Consequently, the behavior of the wave function components in the latter two channels is quite different than for the other six.
First, the spin-exchange interaction $bb\rightarrow ac\rightarrow bb$ is relatively strong, suppressed by only a small energy denominator.  Furthermore, the $bb$ and $ac$ components can persist to much larger distances than the other six, extending far beyond the van der Waals length (as with a halo dimer) while the remaining six are short ranged, i.e. their amplitude decays quickly outside the van der Waals length $R_\text{vdW}$.
Since $R_\text{vdW}$ is the smallest relevant length scale for our problem (we can ignore short range ``chemistry''), it can be thought of as vanishingly small.
A diatomic channel that is confined within this range thus resembles a point-like molecule.
As is typical for two-channel models~\cite{Petrov04,Gurarie07}, also here, we categorize the six channels with components confined to short range (order $R_\text{vdW}$ or less) as ``molecular channels'' while the two channels with components that extend far beyond $R_\text{vdW}$ are called  ``atomic channel''.
We will use this nomenclature throughout this paper.

In the model in Section~\ref{sec:model + hamiltonian} below for the $bb$-channel we consider both long-range components as separate atomic channels and group all closed channels together~\cite{bb_channel_meaning}, thus reducing an eight-channel problem to an effective three-channel model.
Utilizing this reduction, the magnetic field dependence of the populations is shown in Fig.~\ref{fig:Li7:wave functions and population}(b) and discussed below.

In contrast to the eight-channel $bb$ case, the $aa$-channel, which features an isolated Feshbach resonance at $B_{\text{res},aa}=737.69(2)$~G~\cite{Julienne14}, has no near-lying channel with the same spin-projection, $M_{\text{tot}}=2$.
Hence, all additional four channels ($ag$, $bh$, $fh$ and $gg$) collectively form the molecular channel, thus making the $aa$-channel a good example of the reduction of a homo-nuclear five-channel case to an effective two-channel model, as described in Ref.~\cite{Mies00}.

We note that, within the coupled channels framework, the reduction from eight (five) to three (two) channels is more complicated than just grouping together the "molecular channels."
In reality, an ``atomic channel'' can also have a non-zero contribution to the molecular bound state at short range~\cite{Mies00}.

Comparing the scattering lengths and dimer binding energies of the $aa$ and $bb$ channels demonstrates the effect of a second, near-degenerate continuum.
To this end we use full coupled-channels calculations~\cite{ccOnlySWave} and shift the observables of both channels by their respective resonance position.

It is worth mentioning that the two resonances of the $bb$ channel and the resonance of the $aa$ channel all arise due to different nuclear spin projections of the same bare molecular $\nu=41$, $J=0$ bound state in the singlet potential~\cite{Julienne14}.
In addition, since all three resonances are at large magnetic fields, the free atom states have both electron spins down and thus correspond to molecular triplet states.
Thus, the differential magnetic moments of the bare molecule in the $bb$ and $aa$ channels are almost the same (within $3\%$).

\begin{figure}
\includegraphics[width=\linewidth]{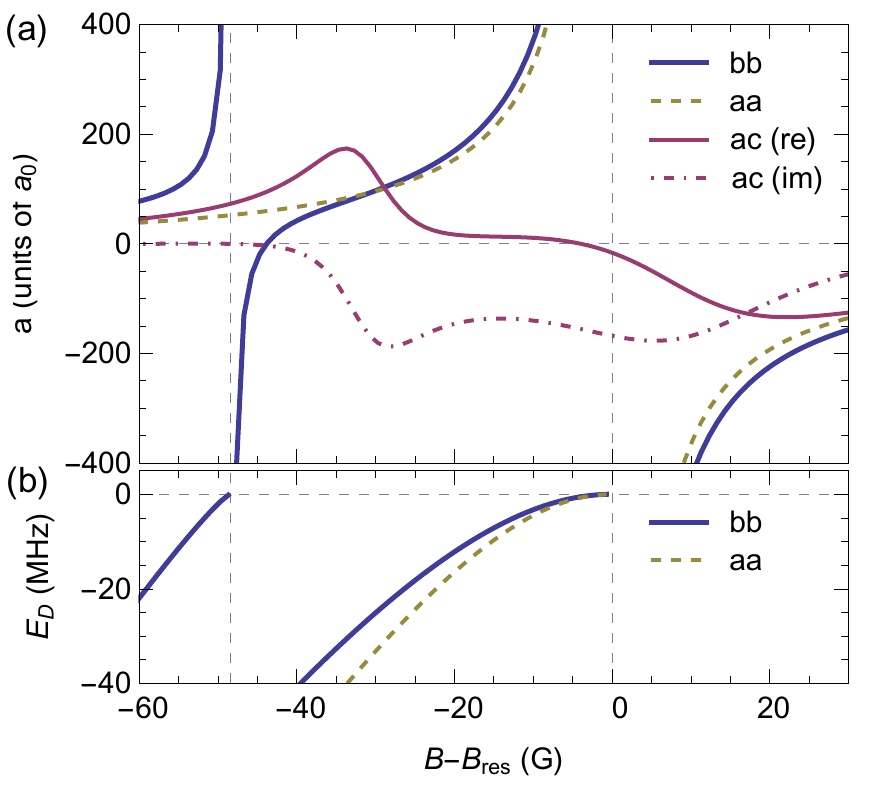}
\caption{\label{fig:Li7:scat and dimer}
The magnetic field dependence of the $^7$Li $bb$- and $aa$-channel (a) scattering lengths and (b) dimer binding energies.
}
\end{figure}

The scattering lengths of the $aa$ and $bb$ channels, shown in Fig.~\ref{fig:Li7:scat and dimer}(a), are overwhelmingly similar.
The main difference between the two channels is caused by the additional resonance in the $bb$-channel.
Its existence forces a zero-crossing, and thus a large gradient, at $B-B_{\text{res},bb}=-44$ G which is absent in the $aa$-channel.
We also show the $ac$-scattering length, which is complex because of inelastic collisional loss to the open $bb$-continuum due to strong spin-exchange coupling.
Its structure, in particular the double peak of the imaginary part and the inflection point of the real part, are due to the overlapping nature of the two resonances.

The similarity of the $bb$ and $aa$ scattering lengths suggests that also the two dimer binding energies are comparable.
Moreover, the additional dimer associated with the narrow resonance in the $bb$-channel is expected to make the $bb$ dimer near $893.7$ G more deeply bound due to dimer-dimer level repulsion (see Ref.~\cite{Yudkin21}, in particular Appendix C therein).
However, as seen in Fig.~\ref{fig:Li7:scat and dimer}(b), the $bb$-channel dimer is considerably shallower than the $aa$-channel dimer and has a somewhat larger universal range where $E_D\sim a^{-2}$~\cite{Li7_rescaleAxis}.

The populations $0\leq\int\left|\phi_{ij}\left(R\right)\right|^2dR\leq1$ of the various channels are shown in Fig.~\ref{fig:Li7:wave functions and population}(b).
While the atomic $bb$ and $aa$ channels have similar behavior, the molecular channels are different.
In the $bb$-case, the molecular channel (which dominates far from resonance, i.e. at deep dimer binding energies) increases at a slower rate due to the population in the $ac$-channel.
The latter reaches a maximal population of $\sim14\;\%$ at $B-B_{\text{res},bb}=-30$ G and then decays like an atomic channel.

The magnetic field dependence of the spin composition of the $aa$-dimer is typical for a closed channel dominated resonance and reproduced by a simple two-channel model with one open atomic and one closed molecular channel~\cite{Petrov04,Castin06,Petrov12}; see also Ref.~\cite{Mies00}.
In contrast, the unique behavior of the $ac$ population marks the appearance of a non-trivial spin composition which we successfully model with a three-channel model ($bb$, $ac$ and a closed molecular channel) in the following.
The naturally arising question then concerns the three-body sector:
How is the Efimov trimer affected by the proximity of the $ac$-channel and its non-negligible population?
Finding an answer within a full coupled-channels model is extremely resource intensive.
The straight-forward and comprehensive three-channel model introduced here and benchmarked in the two-body sector can be applied to the three-body sector without major complications.
In addition, due to the model's simplicity, any deviation from the two-channel model is necessarily caused by the $ac$-channel.

\section{Model Description}
\label{sec:model + hamiltonian}

We consider spin-$1$ particles (i.e. total spin $F=1$) with spin projections $m_F=1,0,-1$ which we label $\sigma=a,b,c=1,2,3$ respectively (lowest three states in Fig.~\ref{fig:Li7:Zeeman shift}).
In contrast to other studies of few-body physics in a spinor gas~\cite{Collusi14,Mestrom21}, we work at large magnetic fields ($\mu_BB\gg A_{hf}$) and with a single, well-defined, entrance spin channel (the $bb$-channel).
The model is designed to capture resonant scattering and associated loosely bound states.
To this end, we assume, in addition to atomic channels formed from $\sigma=a,b,c$, the presence of diatomic molecular levels with energy detuning $E_{\text{mol},\sigma}$.
In cold-atom experiments, $E_{\text{mol},\sigma}$ can be tuned via an external magnetic field.
For the closed molecular channels, the index $\sigma$ indicates the spin flavor not part of the molecule, and has a spin-projection of, e.g., $M^{\left(\sigma=3\right)}_F=m^{\left(\sigma=1\right)}_F+m^{\left(\sigma=2\right)}_F$.
We are interested in the zero total-spin projection ($M_{\text{tot}}=0$) two- and three-body systems made up of these particles.
Thus, in the two-body sector there are two atomic channels (continua), $bb$ and $ac$, which are coupled via a single $M_F=0$ molecular state (closed channel).
For the energies of interest, the atomic $bb$-($ac$-)channel is open (closed).
The three-body sector also features two continua, namely $bbb$ (open) and $abc$ (closed).
However, a three-body bound state with $M_{\text{tot}}=0$ can be composed from a diatomic molecule and a supplementary particle in three different ways: (1) a $M_F=0$ molecule with a $m_F=0$ atom, (2) a $M_F=+1$ molecule with a $m_F=-1$ atom, and (3) a $M_F=-1$ molecule with a $m_F=1$ atom.

We proceed by writing the Hamiltonian of the system:
\begin{equation}
\hat{H}=\hat{H}_{\text{atom}}+\hat{H}_{\text{mol}}+\hat{H}_{\text{int}}.
\label{eq:Hamiltonian:full}
\end{equation}
The atomic and molecular parts, which are the sum of all three spin flavours: $\hat{H}_{\text{atom/mol}}=\sum_{\sigma}\hat{H}_{\text{atom/mol},\sigma}$, include kinetic energy and an energy detuning:
\begin{equation}
\hat{H}_{\text{atom},\sigma}=\int\frac{d^{3}k}{\left(2\pi\right)^{3}}\left(\frac{\hbar^{2}k^{2}}{2m}+E_{\sigma}\right)\hat{a}_{\sigma,\vec{k}}^{\dagger}\hat{a}_{\sigma,\vec{k}}
\end{equation}
\begin{equation}
\hat{H}_{\text{mol},\sigma}=\int\frac{d^{3}k}{\left(2\pi\right)^{3}}\left(\frac{\hbar^{2}k^{2}}{4m}+E_{\text{mol},\sigma}\right)\hat{d}_{\sigma,\vec{k}}^{\dagger}\hat{d}_{\sigma,\vec{k}}.
\end{equation}
Here, $\hat{a}_{\sigma,\vec{k}}^{\dagger}$ ($\hat{d}_{\sigma,\vec{k}}^{\dagger}$) creates an atom (a molecule) of type $\sigma$, mass $m$ ($2m$) and with energy detuning $E_\sigma$ ($E_{\text{mol},\sigma}$) and momentum $\hbar\vec{k}$.
The molecular detuning $E_{\text{mol},\sigma}$ is tunable and serves as a surrogate for the magnetic field detuning from resonance.
The interaction has two parts:
\begin{equation}
\hat{H}_{\text{int}}=\sum_{\sigma}\hat{H}_{\text{int},\sigma}+\hat{H}_{\text{int},22}.
\end{equation}
The first takes two atoms different from $\sigma$ and turns them into a molecule of type $\sigma$ or vice-versa.
Written explicitly for $\sigma=3$ it is
\begin{multline}
\hat{H}_{\text{int},3}=\Lambda_{3}\int\frac{d^{3}q}{\left(2\pi\right)^{3}}\int\frac{d^{3}k}{\left(2\pi\right)^{3}} \\
\left(\hat{d}_{3,\vec{q}}^{\dagger}\hat{a}_{1,\vec{k}+\frac{\vec{q}}{2}}\hat{a}_{2,-\vec{k}+\frac{\vec{q}}{2}}+\hat{a}_{2,-\vec{k}+\frac{\vec{q}}{2}}^{\dagger}\hat{a}_{1,\vec{k}+\frac{\vec{q}}{2}}^{\dagger}\hat{d}_{3,\vec{q}}\right).
\end{multline}
The second part takes two $\sigma=2$ atoms to create a $\sigma=2$ molecule:
\begin{multline}
\hat{H}_{\text{int},22}=\Lambda_{22}\int\frac{d^{3}q}{\left(2\pi\right)^{3}}\int\frac{d^{3}k}{\left(2\pi\right)^{3}} \\
\left(\hat{d}_{2,\vec{q}}^{\dagger}\hat{a}_{2,\vec{k}+\frac{\vec{q}}{2}}\hat{a}_{2,-\vec{k}+\frac{\vec{q}}{2}}+\hat{a}_{2,-\vec{k}+\frac{\vec{q}}{2}}^{\dagger}\hat{a}_{2,\vec{k}+\frac{\vec{q}}{2}}^{\dagger}\hat{d}_{2,\vec{q}}\right).
\end{multline}
We stress that the molecule participating in $\hat{H}_{\text{int},22}$ is the same as the one in $\hat{H}_{\text{int},2}$, namely a molecule with $M_F=0$ spin projection.

The Hamiltonian in Eq.~(\ref{eq:Hamiltonian:full}) does not include scattering within the $bb$- or $ac$-channel, given that they are often, and in particular for $^7$Li, negligible with respect to the resonant scattering considered here.
An exception occurs if a zero-energy resonance is present, see e.g. Ref.~\cite{Arndt97,Abraham97}.
Further, direct $bb$-to-$ac$ coupling is also not included.
A spin-exchange of this sort is possible indirectly via the $M_F=0$ molecule and therefore resonantly enhanced around the Feshbach resonance.
The direct coupling is of the same form as background scattering within either of the atomic channels and neglected here for the sake of simplicity.

The Hamiltonian can be viewed as the amalgamation of a homo- and hetero-nuclear version of the two-channel model, where the synthesis is made by merging the homo-nuclear molecule with the same-spin hetero-nuclear molecule.
For $\Lambda_\sigma=0$ ($\Lambda_{22}=0$) the standard homo-(hetero-)nuclear two-channel model is reobtained, allowing quantitative comparisons in what follows.

\section{Two-Body}
\label{sec:2-body}

\begin{figure}[b]
\includegraphics[width=\linewidth]{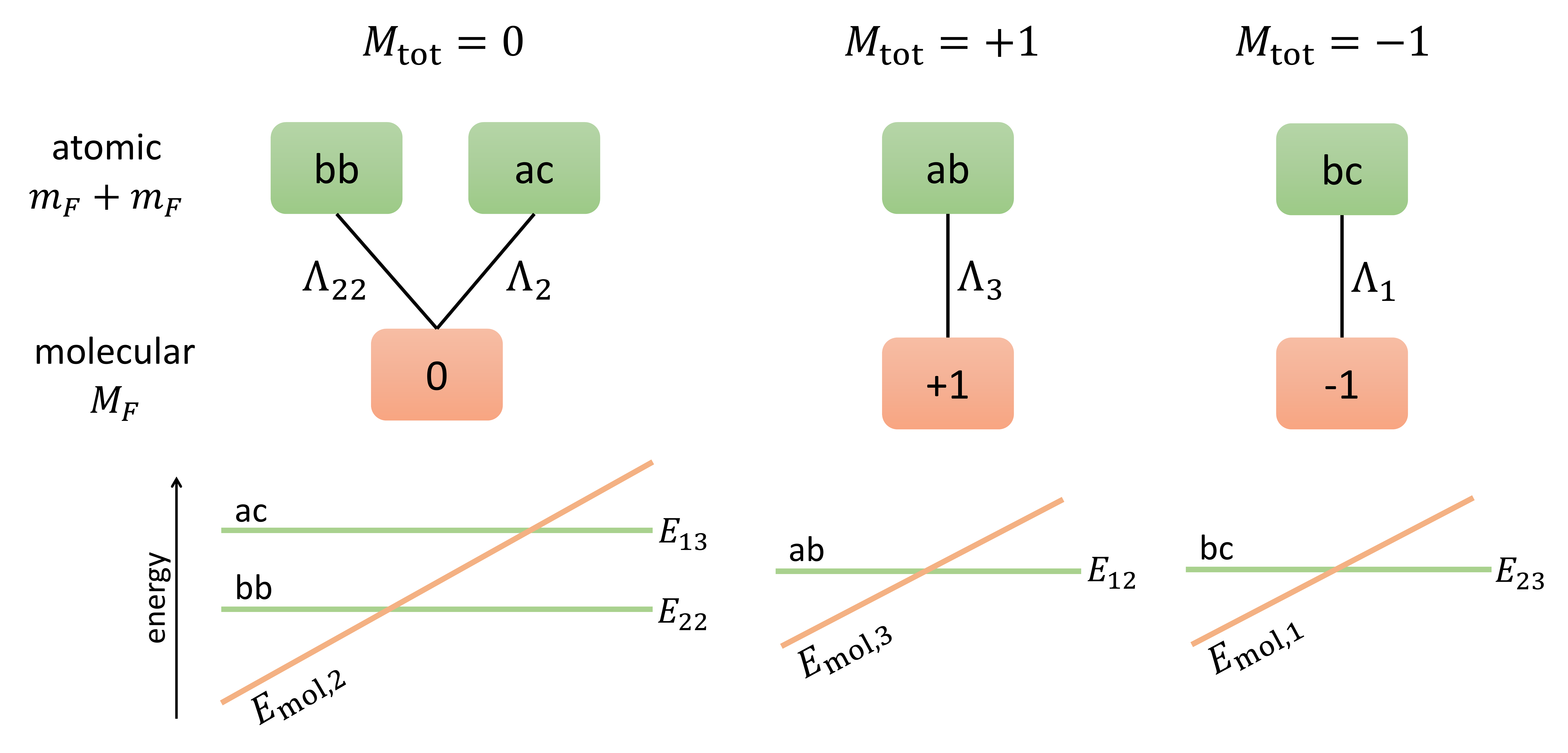}
\caption{\label{fig:2body:connectivity}
Depiction of the three possible two-body sectors in $F=1$.
The upper illustration shows the bare states and their coupling via $\hat{H}_{\text{int}}$.
Below, the bare state energy-dependence on the parameter controlling $E_{\text{mol},\sigma}$ is illustrated.}
\end{figure}

In the $M_{\text{tot}}=0$ two-body sector, the most general two-body wave function is
\begin{multline}
|\psi_{0}\rangle=\beta_{2}\hat{d}_{2,\vec{q}=0}^{\dagger}|0\rangle \\
+\int\frac{d^{3}k}{\left(2\pi\right)^{3}}\alpha_{ac}\left(\vec{k}\right)\hat{a}_{3,\vec{k}}^{\dagger}\hat{a}_{1,-\vec{k}}^{\dagger}|0\rangle \\
+\int\frac{d^{3}k}{\left(2\pi\right)^{3}}\alpha_{bb}\left(\vec{k}\right)\hat{a}_{2,\vec{k}}^{\dagger}\hat{a}_{2,-\vec{k}}^{\dagger}|0\rangle
\label{eq:2body:wave function}
\end{multline}
where we have chosen the center-of-mass reference frame.
As is the case in $^7$Li we choose the energy $E_{22}=2E_2$ of the $bb$-channel to be lower than the energy $E_{13}=E_1+E_3$ of the $ac$-channel: $E_{13}>E_{22}$.
The three bare states are illustrated in Fig.~\ref{fig:2body:connectivity} together with the couplings induced by the Hamiltonian.
The decoupled homo-(hetero-)nuclear wave function $|\psi_{\text{hom}}\rangle$ ($|\psi_{\text{het}}\rangle$) is given by plugging in $\alpha_{ac}=0$ ($\alpha_{bb}=0$) (App.~\ref{app:hetero}).
In the $M_{\text{tot}}=\pm1$ two-body sectors, $|\psi_{+1}\rangle$ and $|\psi_{-1}\rangle$ are equivalent to $|\psi_{\text{het}}\rangle$.

In the following we solve the Schr\"odinger equation $\hat{H}|\psi_{0}\rangle=E|\psi_{0}\rangle$ and obtain expressions for the scattering properties and the associated bound state.
Direct substitution of $\hat{H}$ and $|\psi_{0}\rangle$ into $\hat{H}|\psi_{0}\rangle=E|\psi_{0}\rangle$ leads to three coupled equations, one for each bare state projection.
The free particle amplitudes are of Lorentzian form [see Eq.~(\ref{eq:2body:atomic amplitudes}) below] and thus a three-dimensional integral over them diverges.
In order to avoid infinities, a high-momentum cut-off $k_{\text{cut-off}}$ must be introduced, which we use to renormalize all of the quantities appearing in $\hat{H}$ and $|\psi_{0}\rangle$, such that they become dimensionless.
For example, the renormalized momentum $k$ is $\tilde{k}=k/k_{\text{cut-off}}$, the renormalized molecular amplitude $\beta_2$ is $\tilde{\beta}_2=\beta_2k_{\text{cut-off}}^{3/2}$ and the renormalized coupling constants $\Lambda_{\sigma,22}$ are $\tilde{\Lambda}_{\sigma,22}=\Lambda_{\sigma,22}k_{\text{cut-off}}^{3/2}/E_{\text{cut-off}}$, where $E_{\text{cut-off}}=\hbar^2k_{\text{cut-off}}^2/m$ is the cut-off energy.
Note that the atomic amplitudes $\alpha_{\text{ac,bb}}$ are already dimensionless and hence do not need renormalizing.
After this step, $k_{\text{cut-off}}$ no longer appears in the coupled equations.
Hernceforth, all dimensionless, renormalized quantities are denoted by a tilde.

For demonstration, the various two-body observables are considered using the parameters in Table~\ref{tb:2body:parameters} and we compare them to those of the homo- and hetero-nuclear case.
We choose the same coupling constants for all models, i.e. $\tilde{\Lambda}_{22}=\tilde{\Lambda}_{2}=\tilde{\Lambda}_{\text{hom}}=\tilde{\Lambda}_{\text{het}}$, to keep the microscopic models identical and thus allowing quantitative comparison.
We note that, due to Bose-enhancement, the effective coupling of the homo-nuclear case is $\sqrt{2}$ larger than in the hetero-nuclear case, leading to the former resonance being broader (App.~\ref{app:homo and hetero}).
The comparison between the $M_{\text{tot}}=0$ model and the homo-nuclear model is equivalent to comparing the $^7$Li $bb$- and $aa$-channel observables.

\begin{table}[b]
\centering
\begin{tabular*}{\linewidth}{@{\extracolsep{\fill}} c | c c c c}
\hline\hline
 & $bb$ (22) & $ac$ (2) & hom & het \\
\hline
$\tilde{\Lambda}$ & $1$ & $1$ & $1$ & $1$ \\
$\tilde{E}_{\text{thr}}$ & $0$ & $0.0004$ & $0$ & $0$ \\
\hline
$\tilde{E}_{\text{res}}$ & $0.1504$ & $0.1524$ & $0.1013$ & $0.0507$ \\
$\tilde{R}^\star$ & $18.78$ & - & $6.28$ & $12.57$ \\
\hline\hline
\end{tabular*}
\caption{\label{tb:2body:parameters}
The first two rows list the parameters used for Figs.~\ref{fig:2body:resonance position and r star}-\ref{fig:2body:scat dimer pop}.
The latter two rows are the obtained resonance positions and strengths.}
\end{table}

\subsection{Scattering Properties}

Here, for two atoms in the $bb$-channel with energy $E=E_{22}+E_0$, where $E_0=\hbar^2k_0^2/m>0$ is their relative kinetic energy, we derive an expression for the elastic scattering amplitude $f_{bb}\left(k_0\right)$.
For $E_0<E_{13}$ the $ac$-channel is closed and $bb\rightarrow bb$ is the only possible scattering event.
However for $E_0\geq E_{13}$ an additional scattering process is energetically allowed, namely inelastic scattering $bb\rightarrow ac$.
Hence, the elastic and inelastic scattering processes compete for incoming kinetic energies above the $ac$ threshold and $f_{bb}\left(k_0\right)$ is reduced with respect to its two-channel model counterpart.
Likewise, two atoms in the $ac$-channel with energy $E=E_{13}+E_0$ can either scatter elastically back into the $ac$-channel or inelastically to the $bb$-channel.

\begin{figure}
\includegraphics[width=\linewidth]{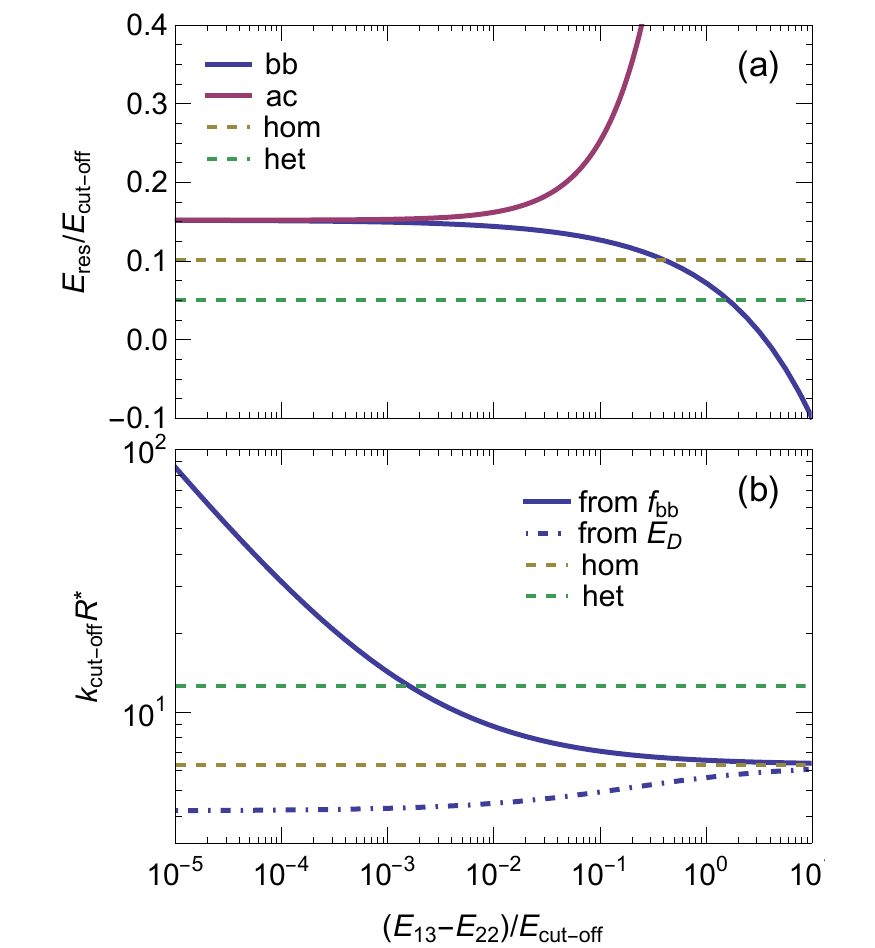}
\caption{\label{fig:2body:resonance position and r star}
$E_{13}$-dependence.
(a) The $bb$- and $ac$-resonance positions are shown as a function of the energy splitting between the continua.
(b) The two possible ways to extract $R^\star_{bb}$ are contrasted.
The solid line is Eq.~(\ref{eq:2body:r star of bb-channel}).
}
\end{figure}

In order to find $f_{bb}\left(k_0\right)$ we use $\hat{H}|\psi_{0}\rangle=\left(E_{22}+E_0\right)|\psi_{0}\rangle$ to write the $bb$ and $ac$ atomic channel amplitudes $\alpha_{bb}\left(\tilde{k}\right)$ and $\alpha_{ac}\left(\tilde{k}\right)$ in terms of the closed molecular amplitude $\tilde{\beta}_2$:
\begin{subequations}
\begin{equation}
\alpha_{bb}\left(\tilde{k}\right)=\left(2\pi\right)^{3}\delta\left(\tilde{k}-\tilde{k}_{0}\right)-\frac{\tilde{\Lambda}_{22}\tilde{\beta}_{2}}{\tilde{k}^{2}-\tilde{k}_{0}^{2}}
\end{equation}
\begin{equation}
\alpha_{ac}\left(\tilde{k}\right)=\left(2\pi\right)^{3}\delta\left(\tilde{k}-\sqrt{\tilde{k}_{0}^{2}-\Delta\tilde{E}}\right)-\frac{\tilde{\Lambda}_{2}\tilde{\beta}_{2}}{\tilde{k}^{2}-\tilde{k}_{0}^{2}+\Delta\tilde{E}}
\end{equation}
\label{eq:2body:atomic amplitudes}
\end{subequations}
where $\Delta\tilde{E}=\tilde{E}_{13}-\tilde{E}_{22}>0$ is the energy difference of the two atomic channels.
From a scattering theory perspective the $\delta$-function represents an incoming plane wave and the second term is the scattered, spherically symmetric, wave.
For the waves in $\alpha_{bb}$ the modulus of the $k$-vector is $\tilde{k}_0$ such that both waves exist for any kinetic energy.
However, in the case of $\alpha_{ac}$ it is $\sqrt{\tilde{k}_{0}^{2}-\Delta\tilde{E}}$, which is real (i.e. the waves propagate) only if the kinetic energy $\tilde{k}_0^2$ is equal of larger than $\Delta\tilde{E}$.
Further, the origin of the scattered wave is twofold.
Although all scattering events must go through the molecule due to the connectivity defined by the Hamiltonian (Fig.~\ref{fig:2body:connectivity}), the molecule can originate from either the $bb$ or the $ac$ atomic channel.
This is made clear by the form of the molecular amplitude, which is given by
\begin{equation}
\tilde{\beta}_{2}\left(\tilde{k}_{0}\right)=-\frac{2\tilde{\Lambda}_{22}\Theta\left(\tilde{k}_{0}\right)+\tilde{\Lambda}_{2}\Theta\left(\tilde{k}_{0}-\sqrt{\Delta\tilde{E}}\right)}{D\left(\tilde{k}_{0}\right)},
\end{equation}
where
\begin{multline}
D\left(\tilde{k}_{0}\right)=\left(\tilde{E}_{\text{mol},2}-\tilde{E}_{22}-\tilde{k}_{0}^{2}\right) \\
-\frac{\tilde{\Lambda}_{2}^{2}}{2\pi^{2}}\left(1-\frac{\pi}{2}\sqrt{\Delta\tilde{E}
-\tilde{k}_{0}^{2}}\right)-\frac{\tilde{\Lambda}_{22}^{2}}{\pi^{2}}\left(1-i\frac{\pi}{2}\tilde{k}_{0}\right).
\end{multline}
The amplitude of the molecular channel is the sum of two terms corresponding to the two origins.
The Heaviside step functions $\Theta\left(x\right)$ arise due to the two different continuum thresholds and prevent an influx from the $ac$-channel if it is closed.
The factor $2$ in the first term is due to Bose-enhancement.
The scattering amplitude for elastic $bb\rightarrow bb$ scattering is thus
\begin{equation}
\tilde{f}_{bb}\left(\tilde{k}_{0}\right)=\frac{\tilde{\Lambda}_{22}^{2}}{2\pi}\frac{\Theta\left(\tilde{k}_{0}\right)}{D\left(\tilde{k}_{0}\right)},
\end{equation}
where we have used the scattered wave from $\alpha_{bb}$ and the molecular amplitude term originating from the $bb$-channel.
To derive an expression for the $ac\rightarrow ac$ elastic scattering amplitude we consider the kinetic energy $\delta E_0=\hbar^2(\delta k_0)^2/m>0$ measured with respect to the $ac$-threshold (the excess energy) and defined via $E=E_{22}+\Delta E+\delta E_0$ or $\tilde{k}_0^2=\Delta\tilde{E}+\delta\tilde{k}_0^2$.
Taking the scattered wave from $\alpha_{ac}$ and the molecular amplitude term originating from the $ac$-channel we find
\begin{equation}
\tilde{f}_{ac}\left(\delta\tilde{k}_{0}\right)=\frac{\tilde{\Lambda}_{2}^{2}}{4\pi}\frac{\Theta\left(\delta\tilde{k}_{0}\right)}{D\left(\sqrt{\Delta\tilde{E}+
\delta\tilde{k}_{0}^2}\right)}.
\end{equation}

\begin{figure}
\includegraphics[width=\linewidth]{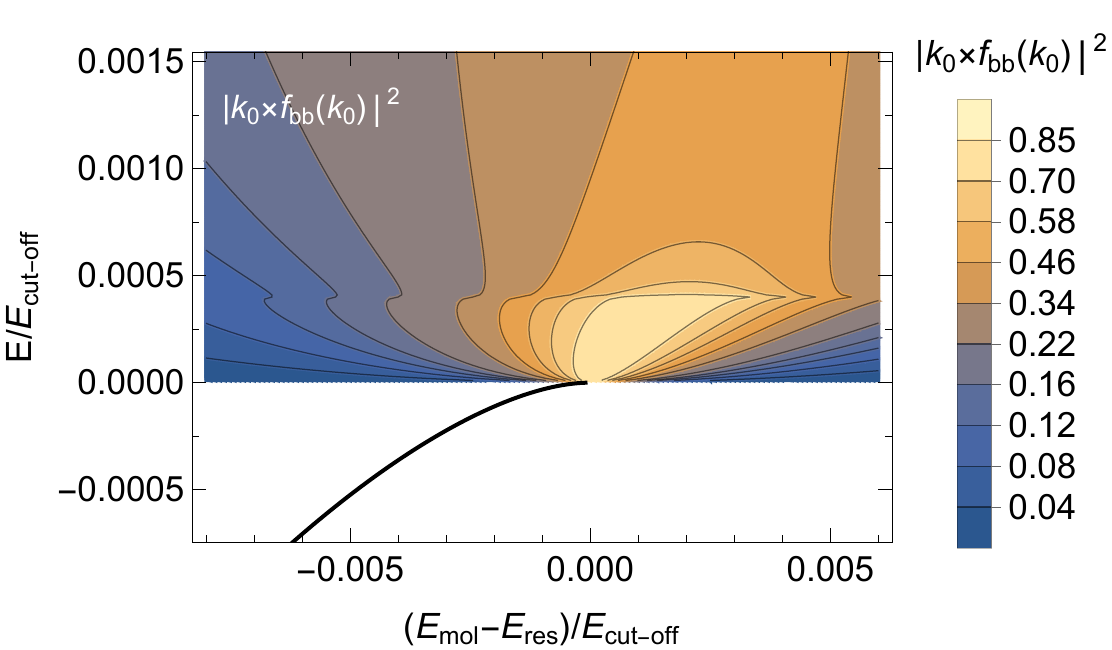}
\caption{\label{fig:2body:scattering amplitude on linear scale}
For $E>0$, a contour plot of $\left|\tilde{k}_0\cdot\tilde{f_{bb}}\left(\tilde{k}_0\right)\right|^2$, as a function of the bare molecular energy $\tilde{E}_{\text{mol}}$ (shifted to resonance) and the kinetic energy $\tilde{E}_0$ of two free particles is shown.
The dimer binding energy, which vanishes at resonance, is depicted for $E<0$.
These $M_{\text{tot}}=0$-model properties are directly comparable to the full coupled-channels calculations of Fig.~\ref{fig:Li7:scattering amplitude}.}
\end{figure}

\begin{figure}
\includegraphics[width=\linewidth]{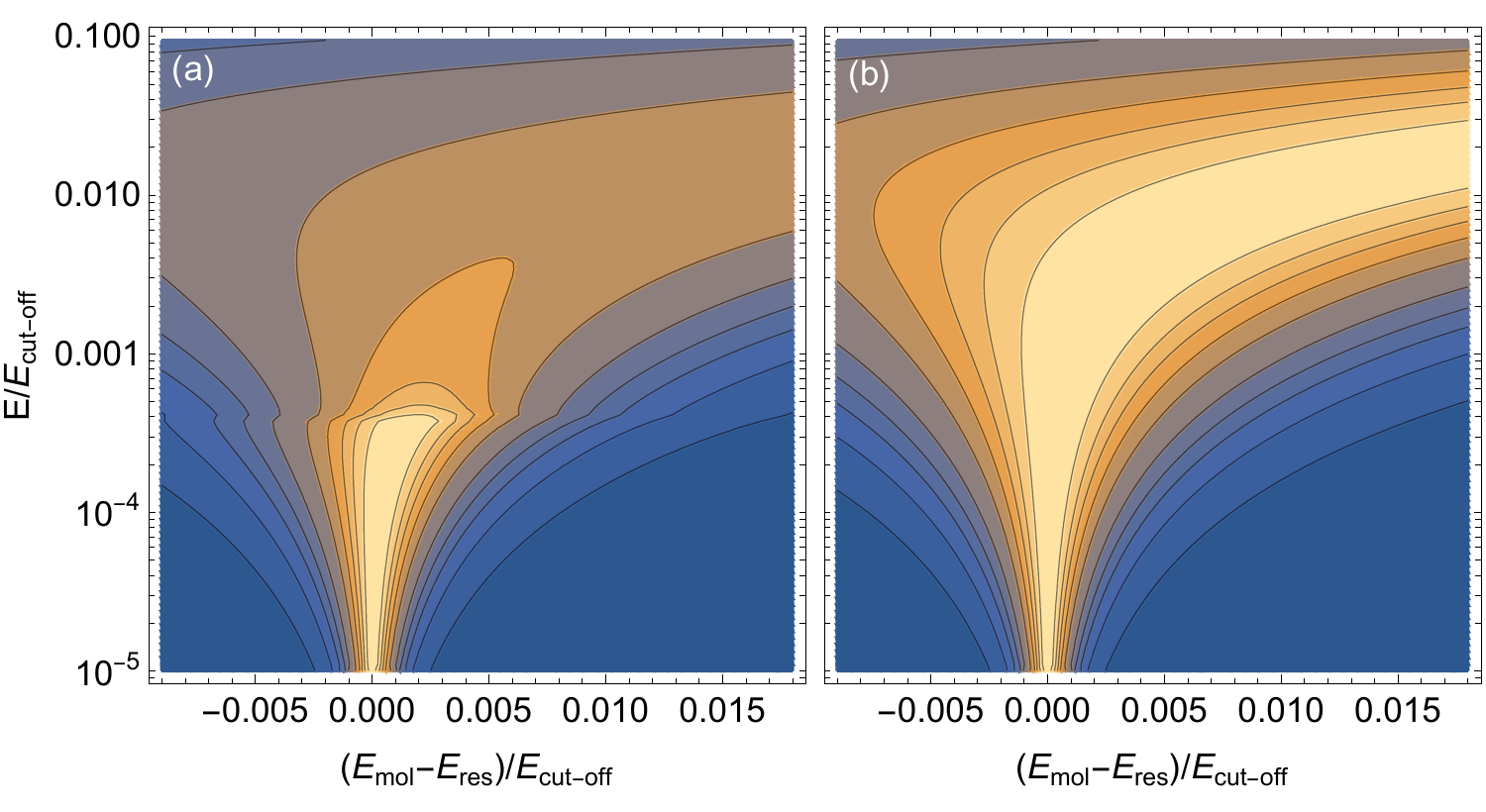}
\caption{\label{fig:2body:scattering amplitude}
(a) Contour plot of $\left|\tilde{k}_0\cdot\tilde{f_{bb}}\left(\tilde{k}_0\right)\right|^2$, as in Fig.~\ref{fig:2body:scattering amplitude on linear scale} but with a logarithmic vertical scale.
(b) For comparison, the two-channel homonuclear case is shown.
The color scales here and in Figs.~\ref{fig:Li7:scattering amplitude} and~\ref{fig:2body:scattering amplitude on linear scale} are identical.}
\end{figure}

As a first observable we extract the resonance position $\tilde{E}_{\text{res},i}$ ($i=\left\{bb,ac,\text{hom},\text{het}\right\}$), i.e. the value the bare molecular energy $\tilde{E}_{\text{mol},2}$ takes when the scattering length diverges and the condition $\tilde{f}_{i}^{-1}\left(\tilde{k}_0=0\right)=0$ is satisfied (Table~\ref{tb:2body:parameters}).
For degenerate continua, i.e. for $E_{13}=E_{22}$, the $bb$- and $ac$-resonance positions are identical and equal to the sum of the homo- and hetero-nuclear model.
For non-equal thresholds the $ac$-resonance position increases linearly with $|E_{13}-E_{22}|$ while the $bb$-resonance decreases as $\sim\sqrt{E_{13}-E_{22}}$; see Fig.~\ref{fig:2body:resonance position and r star}(a).
An energetically distant but equally coupled $ac$-channel ($\tilde{\Lambda}_{22}=\tilde{\Lambda}_2=1$) causes the $bb$-resonance to shift below the homo-nuclear ($\tilde{\Lambda}_2=0$) resonance.
However, realistically $\tilde{\Lambda}_2$ should decrease with increasing $|E_{13}-E_{22}|$.
In the limit $\tilde{\Lambda}_2\rightarrow0$, $\tilde{E}_{\text{res},bb}$ approaches the value of $\tilde{E}_{\text{res},\text{hom}}$.

In the following, the resonance position $E_{\text{res},bb}$ is used to shift the molecular energy axis such that the $bb$-resonance is at the origin.

A contour plot of $0\leq\left|\tilde{k}_0\cdot\tilde{f}_{bb}\left(\tilde{k}_0\right)\right|^2\leq1$, which is equivalent to $\sin^2\eta$~\cite{sinEtaIskf}, demonstrates the effect of the $ac$-threshold embedded in the $bb$-continuum (Fig.~\ref{fig:2body:scattering amplitude on linear scale}).
As in $^7$Li (Fig.~\ref{fig:Li7:scattering amplitude}) we observe cusp behavior at the threshold.
Compared to the homo-nuclear two-channel model, the $bb\rightarrow bb$ (elastic) scattering is decreased above the $ac$-threshold due to the increase in $bb\rightarrow ac$ (inelastic) scattering (Fig.~\ref{fig:2body:scattering amplitude}).

\begin{figure}
\includegraphics[width=\linewidth]{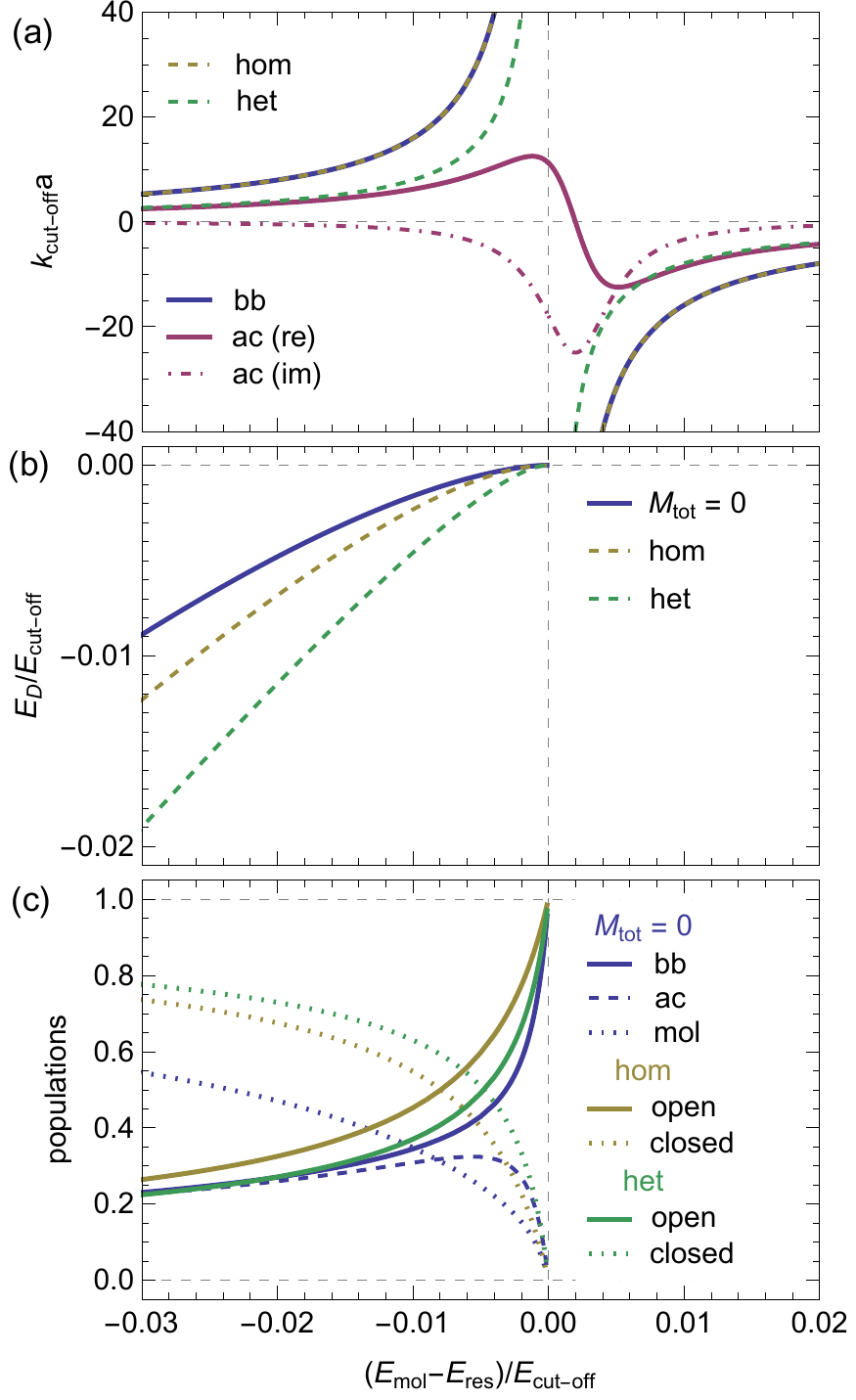}
\caption{\label{fig:2body:scat dimer pop}
Two-body sector.
(a) Plot of the scattering length $\tilde{a}$ as a function of the bare molecular energy $\tilde{E}_{\text{mol}}$.
The $bb$ and $ac$ curves are shifted by $\tilde{E}_{\text{res},bb}$ and hom (het) by $\tilde{E}_{\text{res,hom}}$ ($\tilde{E}_{\text{res,het}}$).
The curves for $bb$ and hom overlap.
(b) Plot of the dimer binding energy with respect to the (lower) continuum threshold.
(c) For the three models in (b) the open and closed channels populations are contrasted.
}
\end{figure}

While the contour plots of Figs.~\ref{fig:2body:scattering amplitude on linear scale} and~\ref{fig:2body:scattering amplitude} are instructive, cold-atom experiments usually operate in the $k_0\rightarrow0$ limit.
In this case one may expand the scattering amplitude in powers of $k_0$ and compare its coefficients to the effective range expansion $\left[\tilde{f}_{bb}\left(\tilde{k}_{0}\right)\right]^{-1}=-\tilde{a}_{bb}^{-1}-i\tilde{k}_0+\tilde{r}_{e,bb}\tilde{k}_0^2/2$, thus obtaining expressions for the scattering length $\tilde{a}_{bb}$ [Fig.~\ref{fig:2body:scat dimer pop}(a)] and effective range $r_{e,bb}$.
The latter is related to the parameter $R^\star_{bb}=-r_{e,bb}/2$ which is indicative of the narrowness of a Feshbach resonance (Table~\ref{tb:2body:parameters}).
In Fig.~\ref{fig:2body:scat dimer pop}(a) we see that $a_{bb}$ is identical to $a_{\text{hom}}$, i.e. its width in units of $E_{\text{mol}}$ is governed by the $bb$-channel.
However, the seemingly perfect overlap is misleading because, in Fig.~\ref{fig:2body:scat dimer pop}(a), we shifted both resonances by their respective (different, see Table~\ref{tb:2body:parameters}) $E_{\text{res},i}$.
More importantly, $R^\star_{bb}>R^\star_{\text{hom}}$, indicating that the $bb$ resonance pole strength is increased by the proximity of the additional $ac$-channel, even beyond $R^\star_{\text{het}}$.
In fact,
\begin{equation}
\tilde{R}_{bb}^{\star}=\frac{2\pi}{\tilde{\Lambda}_{22}^{2}}+\frac{\tilde{\Lambda}_{2}^{2}}{4\tilde{\Lambda}_{22}^{2}}\frac{1}{\sqrt{\tilde{E}_{13}-\tilde{E}_{22}}}
\label{eq:2body:r star of bb-channel}
\end{equation}
is made up of two terms.
The first is the contribution from the $bb$-channel and identical to $\tilde{R}_{\text{hom}}^{\star}$.
The second term is inversely proportional to the square-root of the energy difference.
Thus, the closer the two continua, the stronger the pole.
We note that in the limit $\tilde{\Lambda}_2\rightarrow0$, as well as in the limit $\tilde{E}_{13}\gg\tilde{E}_{22}$, $\tilde{R}_{bb}^{\star}\rightarrow\tilde{R}_{\text{hom}}^{\star}$ as expected.

Turning to the $ac$-threshold, the scattering length $a_{ac}$, which is obtained from $\tilde{a}_{ac}=-\tilde{f}_{ac}\left(\delta\tilde{k}_0=0\right)$, is complex [Fig.~\ref{fig:2body:scat dimer pop}(a)].
At the pole position, $\text{Re}\;a_{ac}=0$ vanishes and $\text{Im}\;a_{ac}$ experiences a maximum.
Expanding $f_{ac}\left(k_0\right)$ beyond the zeroth order is meaningless because the linear term $\sim k_0$ is not parameter independent as required by the optical theorem.

\subsection{Binding Energy}

Now we discuss the negative energy solution of the Schr\"odinger equation, i.e. the binding energy of the Feshbach dimer.
One way of finding it is by looking for poles of the scattering amplitude for $k_0=i\lambda$, $\lambda>0$.
Alternatively, one may solve the Schr\"odinger equation with $E_D=E_{22}-\hbar^2\lambda^2/m<E_{22}$ which reduces to solving the following equation:
\begin{multline}
\left(\tilde{E}_{\text{mol},2}-\tilde{E}_{22}+\tilde{\lambda}^{2}\right)-\frac{\tilde{\Lambda}_{2}^{2}}{2\pi^{2}}\left[1-\frac{\pi}{2}\sqrt{\tilde{E}_{13}-\tilde{E}_{22}+\tilde{\lambda}^{2}}\right] \\
-\frac{\tilde{\Lambda}_{22}^{2}}{\pi^{2}}\left[1-\frac{\pi}{2}\tilde{\lambda}\right]=0.
\label{eq:2body:dimer bb+ac}
\end{multline}
We note that solving the Schr\"odinger equation for $E_D=E_{13}-\hbar^2\lambda^2/m$ leads to a different solution for $\lambda$ but the same (observable) binding energy $E_D$.
The additional atomic channel does {\it not} lead to an additional dimer but instead gives the dimer a mixed $bb$ and $ac$ spin composition.
(This should be contrasted to the three-channel model with two molecular channels~\cite{Yudkin21}.)
Its numerical solution is compared to the homo- and hetero-nuclear dimers in Fig.~\ref{fig:2body:scat dimer pop}(b).
Interestingly, although we established above that the presence of the $ac$-channel increases the pole strength (larger $R^\star$), the $M_{\text{tot}}=0$ dimer is shallower than in the homo-nuclear model and its universal range, where $E_D\sim a^{-2}$, is increased.
In fact, the functional form of the homo-nuclear dimer describes the $M_{\text{tot}}=0$ dimer with $\tilde{\Lambda}=1.12$ or $\tilde{R}^\star=4.364$.
We conclude that, counter intuitively, the addition of the $ac$-channel increases the pole strength while also increasing the universal range, a correlation that is usually vice-versa.

To illustrate this point in a more general setting, Fig.~\ref{fig:2body:resonance position and r star}(b) contrasts the value of $\tilde{R}^\star_{bb}$ found by expanding the scattering amplitude [Eq.~(\ref{eq:2body:r star of bb-channel})] to the value of $\tilde{R}^\star_{bb}$ obtained from fitting the $M_{\text{tot}}=0$ dimer to the homo-nuclear dimer.
While they become equal at large continua separation, as they approach degeneracy the two values grow apart.
We note that $\tilde{R}^\star_{bb}$ extracted from the dimer approaches $\tilde{R}^\star_{\text{hom}}\tilde{R}^\star_{\text{het}}/(\tilde{R}^\star_{\text{hom}}+\tilde{R}^\star_{\text{het}})$ for $E_{13}\rightarrow E_{22}$ (App.~\ref{app:limit of rStar}).

Some insight into this counter intuitive behaviour may be gained by looking at the populations in the atomic and molecular channels.
Given the dimer binding wave number $\lambda>0$, they are (App.~\ref{app:populations})
\begin{subequations}
\begin{equation}
P_{\text{bb}}=\frac{1}{\mathcal{N}}\left(\frac{\tilde{\Lambda}_{22}^{2}}{8\pi\tilde{\lambda}}\right)
\end{equation}
\begin{equation}
P_{\text{ac}}=\frac{1}{\mathcal{N}}\left(\frac{\tilde{\Lambda}_{2}^{2}}{8\pi\sqrt{\tilde{\lambda}^{2}-\tilde{E}_{22}+\tilde{E}_{13}}}\right)
\end{equation}
\begin{equation}
P_{\text{mol}}=\frac{1}{\mathcal{N}},
\end{equation}
\end{subequations}
where
\begin{equation}
\mathcal{N}=\frac{\tilde{\Lambda}_{22}^{2}}{8\pi\tilde{\lambda}}+\frac{\tilde{\Lambda}_{2}^{2}}{8\pi\sqrt{\tilde{\lambda}^{2}-\tilde{E}_{22}+\tilde{E}_{13}}}+1,
\end{equation}
for the $bb$, $ac$ and molecular channel, respectively.
These functions are plotted in Fig.~\ref{fig:2body:scat dimer pop}(c).
Far away from resonance, the population of the $ac$-channel behaves like an atomic channel and approaches zero, as the dimer adopts a predominantly closed channel character.
Close to resonance, the dimer becomes $bb$-open-channel dominated and the $ac$-channel population also approaches zero.
In between it experiences a maximum as $|E_D-E_{22}|\approx|E_{13}-E_{22}|$.
Compared to the open channel populations of the homo- and hetero-nuclear models, the $bb$ population depletes faster, as expected from a stronger pole (larger $R^\star$).
However, the molecular channel population increases at a slower rate than the closed channel in both the homo- and hetero-nuclear models.
This is in agreement with the dimer's binding energy being shallower.
The missing population is of course found in the $ac$-channel.
In other words, the proximity of the $ac$-channel depletes the $bb$-channel faster causing the $bb$-resonance to seem less open-channel dominated (i.e. narrower, larger $R^\star$).
But from the point-of-view of the dimer the atomic channel population is the sum $P_{\text{bb}}+P_{\text{ac}}$ which implies that the resonance is more open-channel dominated (i.e. larger universal range).
It is this duality that gives rise to the (unusual) correlation between the pole strength and the universal range.

Contrasting the observables of the $M_{\text{tot}}=0$ model to the homo-nuclear model is reminiscent of the comparison of the $bb$- and $aa$-channel in $^7$Li (see Fig.~\ref{fig:Li7:scat and dimer}).
We therefore conclude that the nearby $ac$-channel is responsible for the observed difference.
By fitting the model to the coupled-channels scattering length and $R^\star$ of $^7$Li, the three bare parameters of the model (bare resonance position + two coupling constants) can be found.
However, quantitative agreement between the model and the coupled-channels dimer is limited to small binding energies.
We observe that the model predicts a less deeply bound dimer (compared to coupled-channels) for both channels.
We associate this discrepancy to the van der Waals tail of the real interaction potential that is not captured by the model.
We do, however, faithfully reproduce the shallowing effect due to $ac$-channel.

\section{Three-Body}
\label{sec:3-body}

\begin{figure}[b]
\includegraphics[width=\linewidth]{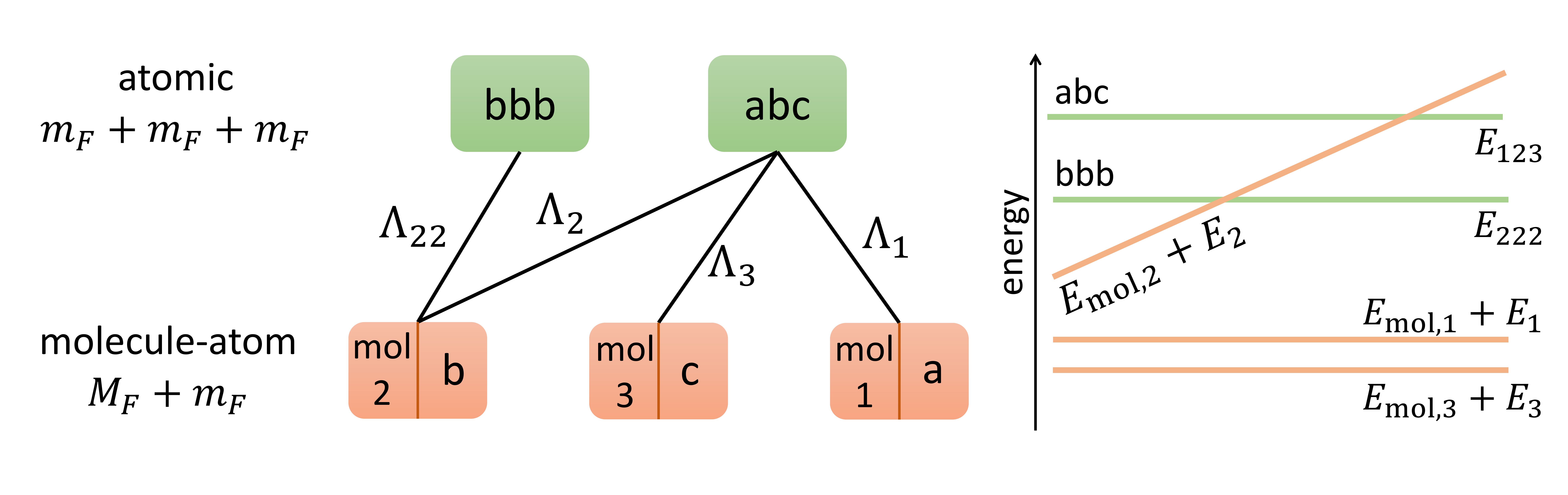}
\caption{\label{fig:3body:connectivity}
Depiction of the $M_{\text{tot}}=0$ three-body sector.
The illustration on the left shows the bare states and their coupling via $\hat{H}_{\text{int}}$.
To the right, the bare state energy-dependence on the parameter controlling $E_{\text{mol},\sigma}$ is illustrated.}
\end{figure}

In a system with three particles and total-spin projection $M_{\text{tot}}=0$ there are two possible spin configurations in the continuum: (i) $bbb$, (ii) $abc$~\cite{3body_aad_channel}.
Each pair ($bb$, $ac$, $ab$, $bc$) is coupled to a molecule with a matching spin projection, and, as discussed above, the $bb$ and $ac$ pairs are coupled via the $M_F=0$ molecule.
The latter is also the link between the two three-body continua (Fig.~\ref{fig:3body:connectivity}).
The most general three-body wave function is
\begin{multline}
|\psi_{\text{3B}}\rangle=\sum_{\sigma}\int\frac{d^{3}q}{\left(2\pi\right)^{3}}\beta_{\sigma}\left(\vec{q}\right)\hat{d}_{\sigma,\vec{q}}^{\dagger}\hat{a}_{\sigma,-\vec{q}}^{\dagger}|0\rangle \\
+\int\frac{d^{3}q}{\left(2\pi\right)^{3}}\int\frac{d^{3}k}{\left(2\pi\right)^{3}}\alpha_{bbb}\left(\vec{q},\vec{k}\right)\hat{a}_{2,\vec{k}+\frac{\vec{q}}{2}}^{\dagger}\hat{a}_{2,-\vec{k}+\frac{\vec{q}}{2}}^{\dagger}\hat{a}_{2,-\vec{q}}^{\dagger}|0\rangle \\
+\int\frac{d^{3}q}{\left(2\pi\right)^{3}}\int\frac{d^{3}k}{\left(2\pi\right)^{3}}\alpha_{abc}\left(\vec{q},\vec{k}\right)\hat{a}_{3,\vec{k}+\frac{\vec{q}}{2}}^{\dagger}\hat{a}_{1,-\vec{k}+\frac{\vec{q}}{2}}^{\dagger}\hat{a}_{2,-\vec{q}}^{\dagger}|0\rangle
\label{eq:3body:wave function}
\end{multline}
There are three decoupled two-body subsystems underlying $|\psi_{\text{3B}}\rangle$, namely $|\psi_{0}\rangle$ [Eq.~(\ref{eq:2body:wave function})], $|\psi_{+1}\rangle$ and  $|\psi_{-1}\rangle$ (Fig.~\ref{fig:2body:connectivity}).
In the three-body sector, these decoupled two-body systems become coupled due to the common continuum (App.~\ref{app:from 2 to 3 body}).

\begin{figure}
\includegraphics[width=\linewidth]{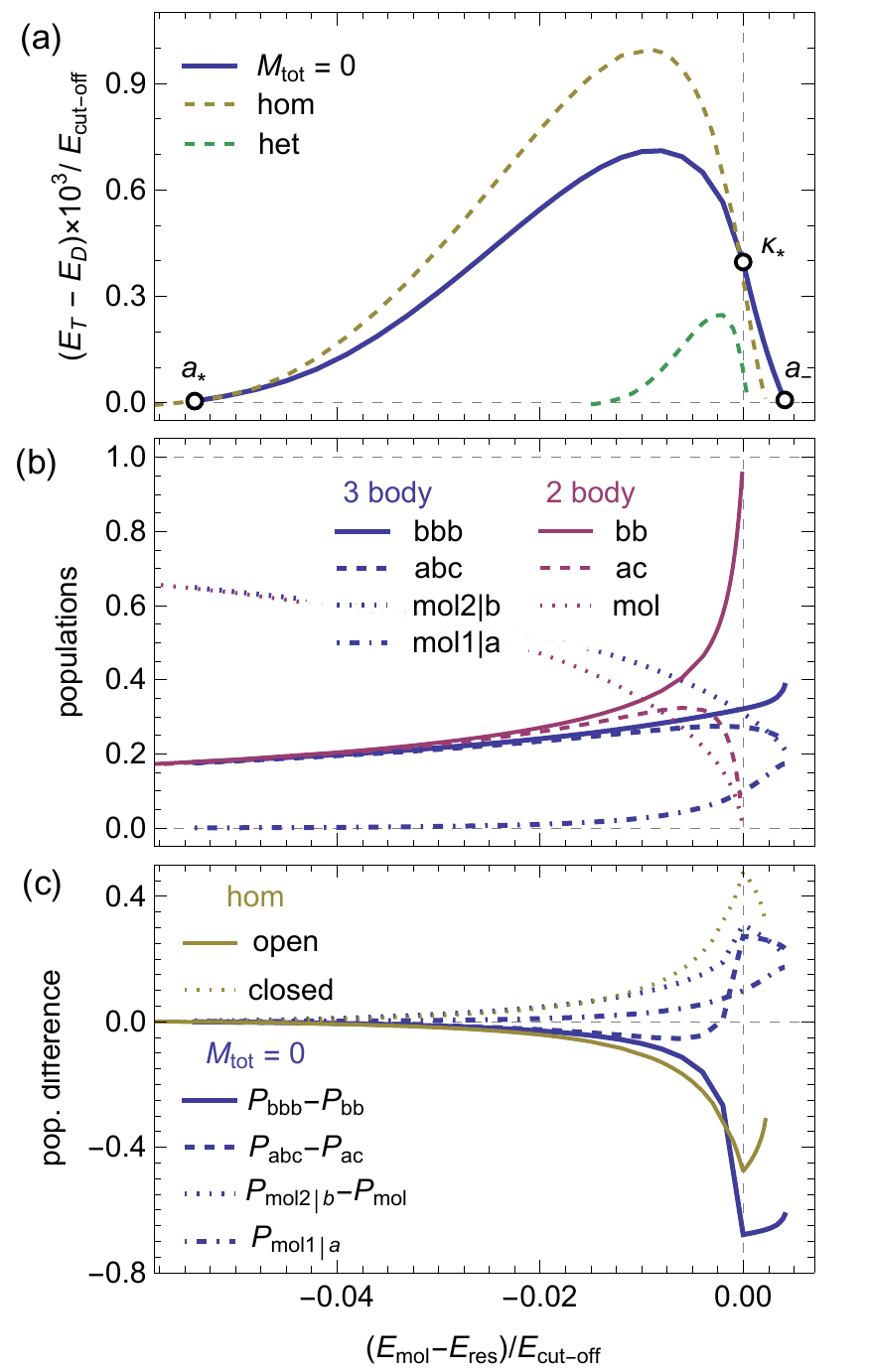}
\caption{\label{fig:3body:trimer + population}
Three-body sector.
(a) Plot of the ground state trimer binding energy with respect to the dimer, as a function of the bare $M_F=0$ molecule energy.
The large difference between the homo- and hetero-nuclear trimer is fully explained by Bose-enhancement (App.~\ref{app:homo and hetero}).
The positions from which the Efimov features in Table~\ref{tb:3body:parameters + features} are extracted are indicated.
(b) The populations in the various open and closed channels are shown for the two- and three-body sector of the $bb$+$ac$ model.
(c) The difference between the three- and two-body populations are compared to those of the homo-nuclear model.}
\end{figure}

Upon solving $\hat{H}|\psi_{\text{3B}}\rangle=E|\psi_{\text{3B}}\rangle$ with $E=E_{222}-\hbar^2\lambda^2/m$ one obtains three coupled equations for the molecule-atom amplitudes $\beta_\sigma\left(q\right)$ and the binding wave number $\lambda$:
\begin{multline}
\left[f_{\sigma}\left(\tilde{q}\right)-g_{\sigma}\left(\tilde{q}\right)\right]\tilde{\beta}_{\sigma}\left(\tilde{q}\right)-\delta_{\sigma,2}g_{22}\left(\tilde{q}\right)\tilde{\beta}_{2}\left(\tilde{q}\right) \\
-\sum_{\sigma^{\prime}\neq\sigma}\int_{0}^{\infty}d\tilde{k}L_{\sigma,\sigma^{\prime}}\left(\tilde{q},\tilde{k}\right)\tilde{\beta}_{\sigma^{\prime}}\left(\tilde{k}\right) \\
-\delta_{\sigma,2}\int_{0}^{\infty}d\tilde{k}L_{22}\left(\tilde{q},\tilde{k}\right)\tilde{\beta}_{2}\left(\tilde{k}\right)=0.
\label{eq:3body:trimer coupled equations}
\end{multline}
Here we have defined the functions
\begin{subequations}
\begin{equation}
f_{\sigma}\left(\tilde{q}\right)=\left(\frac{3\tilde{q}^{2}}{4}+\tilde{E}_{\sigma}+\tilde{E}_{\text{mol},\sigma}-\tilde{E}_{222}+\tilde{\lambda}^{2}\right)
\end{equation}
\begin{equation}
g_{\sigma}\left(\tilde{q}\right)=\frac{\tilde{\Lambda}_{1}^{2}}{2\pi^{2}}\left(1-\frac{\pi}{2}\sqrt{\frac{3\tilde{q}^{2}}{4}+\tilde{E}_{123}-\tilde{E}_{222}+\tilde{\lambda}^{2}}\right)
\end{equation}
\begin{equation}
g_{22}\left(\tilde{q}\right)=\frac{\tilde{\Lambda}_{22}^{2}}{\pi^{2}}\left(1-\frac{\pi}{2}\sqrt{\frac{3\tilde{q}^{2}}{4}+\tilde{\lambda}^{2}}\right)
\end{equation}
\begin{multline}
L_{\sigma,\sigma^{\prime}}\left(\tilde{q},\tilde{k}\right)=\frac{\tilde{\Lambda}_{\sigma}\tilde{\Lambda}_{\sigma^{\prime}}}{4\pi^{2}} \\
\times\ln\left(\frac{\tilde{k}^{2}+\tilde{k}\tilde{q}+\tilde{q}^{2}+\tilde{E}_{123}-\tilde{E}_{222}+\tilde{\lambda}^{2}}{\tilde{k}^{2}-\tilde{k}\tilde{q}+\tilde{q}^{2}+\tilde{E}_{123}-\tilde{E}_{222}+\tilde{\lambda}^{2}}\right)
\end{multline}
\begin{equation}
L_{22}\left(\tilde{q},\tilde{k}\right)=\frac{\tilde{\Lambda}_{22}^{2}}{\pi^{2}}\ln\left(\frac{\tilde{k}^{2}+\tilde{k}\tilde{q}+\tilde{q}^{2}+\tilde{\lambda}^{2}}{\tilde{k}^{2}-\tilde{k}\tilde{q}+\tilde{q}^{2}+\tilde{\lambda}^{2}}\right)
\end{equation}
\end{subequations}
To make the numerical computation of Eqs.~(\ref{eq:3body:trimer coupled equations}) more resource effective we set the parameters of $\sigma=1$ and $\sigma=3$ to be equal, i.e. $\beta_1=\beta_3$, $\Lambda_1=\Lambda_3$, $E_1=E_3$ and $E_{\text{mol},1}=E_{\text{mol},3}$.
This way the three equations in~(\ref{eq:3body:trimer coupled equations}) reduce to two (one for $\sigma=2$ and one for $\sigma=1$).
In addition we fix the value of $E_{\text{mol},1}$ and vary only $E_{\text{mol}}=E_{\text{mol},2}$ as we track the binding energy across the Feshbach resonance.
In Fig.~\ref{fig:3body:trimer + population}(a) we show the spectrum of the ground state trimer, using the parameters in Table~\ref{tb:3body:parameters + features}, and we compare it to the homo- and hetero-nuclear model.
In Table~\ref{tb:3body:parameters + features} we also extract the usual features of the spectrum, namely the merging point with the dimer-atom continuum ($a_\star$), the dissociation into the $bbb$ continuum ($a_-$) and the binding wave number at resonance $\kappa_\star$.
The results of the two-body sector are used to obtain scattering length values from $\tilde{E}_{\text{mol}}$.
We further normalize the features with respect to $R^\star$, such that the high-momentum cut-off cancels and quantitative comparison to experiments is made possible.

\begin{table}
\centering
\begin{tabular*}{\linewidth}{@{\extracolsep{\fill}} c | c c c}
\hline\hline
\multicolumn{4}{c}{Parameters} \\
\hline
channel & $bb$ & $ac$ & $ab$/$bc$ \\
$\sigma$  & $22$ & $2$ & $3$/$1$ \\
\hline
$\tilde{\Lambda}$ & $1$ & $1$ & $1$ \\
$\tilde{E}_{\text{thr}}$ & $0$ & $0.0004$ & $0.0002$ \\
$\tilde{E}_{\text{mol}}-\tilde{E}_{\text{res}}$ & vary & - & $-0.0008$ \\
$\tilde{a}$ & vary & - & $100$ \\
\hline\hline
\multicolumn{4}{c}{Features} \\
\hline
 & $M_{\text{tot}}=0$ & hom & het \\
\hline
$\tilde{a}_\star$ & $2.88$ & $2.88$ & $5.77$ \\
$R^\star/a_\star$ & $6.53$ & $2.18$ & $2.18$ \\
\hline
$\tilde{a}_-$ & $-37.89$ & $-69.2$  & $-137.2$ \\
$R^\star/a_-$ & $-0.496$ & $-0.091$ & $-0.091$ \\
\hline
$\tilde{\kappa}_\star$ & $0.0199$ & $0.0187$ & $0.0094$ \\
$R^\star\kappa_\star$ & $0.374$ & $0.118$ & $0.118$ \\
\hline\hline
\end{tabular*}
\caption{\label{tb:3body:parameters + features}
Top:
Parameters used for Fig.~\ref{fig:3body:trimer + population}.
$\tilde{E}_{\text{thr}}$ denotes the continuum threshold energy.
Bottom:
Values of various features of the ground state Efimov trimer -- see Fig.~\ref{fig:3body:trimer + population}(a).
}
\end{table}

The slope near $a_-$ is smaller than in the two-channel models.
This, however, is an artefact of varying only $E_{\text{mol},2}$ while keeping the others constant.
The same phenomenon is also observed in the pure hetero-nuclear model.

When comparing the $M_{\text{tot}}=0$ spectrum with that of the homo-nuclear two-channel model the impact of the $ac$-channel is apparent.
Although the extreme points of the spectra (almost) overlap, the trimer formed under near-degenerate continua is bound less deep at its maximum.
Thus, the functional form of the spectrum is altered by the presence of the $ac$-channel which suggests that care should be taken in the interpretation of some of the experimental results.
For almost two decades, the majority of Efimov related measurements were conducted at the free atom or atom-dimer threshold with the goal of proving (or disproving) universality.
The comparison in Fig.~\ref{fig:3body:trimer + population}(a) teaches us that the extreme points at their respective continuum thresholds might not suffice to characterize the energy level.

The model is presented here in terms of microscopic parameters such as the coupling constants between the continua and the bare molecules.
When dealing with Feshbach resonances in a cold atom experiment one uses observables, such as the resonance width $\Delta B$ and the $R^\star$ parameter, to characterize the observed behaviour.
For this reason, the Efimov features are shown in units of $R^\star$ in Table~\ref{tb:3body:parameters + features}.
We note that this normalization equates the homo- and hetero-nuclear model because the Bose-enhancement cancels.
To demonstrate the consequence of the $ac$-channel consider a given resonance (given $R^\star$).
If there is no $ac$-channel one expects the merger with the dimer-atom continuum to be at $a_{\star,\text{hom}}=R^\star/2.18$.
However, if one is aware of its presence one anticipates $a_{\star}=R^\star/6.53<a_{\star,\text{hom}}$ to be a factor $\sim3$ lower (for the parameters in Table~\ref{tb:3body:parameters + features}).

As in the two-body sector, also here we can find the populations of the various three-atomic and molecule-atom channels to explore the spin composition of the trimer (App.~\ref{app:3body:populations}).
The obtained expressions are accurate in the limit of small population in the $\sigma=1$ and $3$ molecular channel with respect to the $\sigma=2$ population.
This approximation breaks down for $E_{\text{mol}}\geq E_{\text{res}}$, where the latter decreases and approaches $0$ at $a_-$.
In fact, at $a_-$ we expect the $bbb$-channel population to approach unity while all others vanish.
In Fig.~\ref{fig:3body:trimer + population}(b), where the populations are plotted for the ground state trimer together with the two-body dimer populations, this property is not reproduced due to the approximation breaking down.
At $a_\star$, the trimer populations asymptotically approach the dimer populations indicating the merger of the two energy levels -- the trimer dissociates into a dimer and an atom.
This is also shown in Fig.~\ref{fig:3body:trimer + population}(c), where the difference between the three- and two-body populations is plotted.

The three-body $bbb$ (mol$2|b$) population is smaller (larger) than the two-body $bb$ (mol) population in the entire spectrum.
The same is true in the homo-nuclear model.
However, comparison of the $abc$ and $ac$ populations reveals that, as we move from $a_-$ to $a_\star$, the $abc$ population is larger (like a molecular channel), then crosses the $ac$ population and finally approaches it from below (like an atomic channel).
Thus, close to the free atom (dimer-atom) threshold the $abc$-channel behaves like a molecular (atomic) channel.

Compared to the homo-nuclear model, the molecular channel of the $M_{\text{tot}}=0$ model is less populated.
As in the two-body sector, this is in agreement with its binding energy being smaller.

\section{Implications on Li-7 Trimer}
\label{sec:discussion}

Considering the special characteristics of the $bb$-channel of $^7$Li, namely overlapping resonances and near-degenerate continua, this near-degenerate continua model and our previous overlapping-resonances model~\cite{Yudkin21} encapsulate the main contributions to the asymptotic scattering wave function.
Both are considerably better than the regular two-channel model in the two-body sector and both predict slight alternations to the three-body sector.
However, none agree with the unusual features of the trimer.
Neither of these minimal models can explain why the three-body parameter is universal despite the resonance being narrow or how the trimer crosses into the dimer-atom continuum.
We conclude that the $^7$Li trimer energy is majorly affected by the short range details of the interaction potential whose treatment requires more sophisticated theoretical approaches~\cite{D'IncaoPrivate}.

\section{Conclusion}

We have presented a simple and straightforward, but at the same time realistic and successful, model of resonant interactions in the presence of a near-degenerate continuum.
The phenomenology of the two-body sector of $^7$Li is reproduced.
Since the interaction potential of the model is hard-core (the simplest possible), quantitative agreement is limited.
Inclusion of more interaction potential details, such as the van der Waals tail, should improve thereon, however, the physical mechanism causing the behaviour of the various observables is clear.
In particular, we have shown how a narrow resonance can have an enlarged universal range and how the functional form of the Efimov trimer is altered.

An interesting future aspect to explore is the many-body physics.
The extended universal range of the two-body sector could work in favor of exploring universality near a narrow resonance.

A similar configuration was found to be important for the $p$-wave scattering of identical fermions in K-40 polarized in the $b$-state~\cite{Ahmed-Braun21,KokkelmansPrivate} 

\section*{Acknowledgments}

We acknowledge fruitful discussions with J.~P.~D'Incao and S.~J.~J.~M.~F.~Kokkelmans and members of his research group.

This research was supported in part by the Israel Science Foundation (Grant No. 1543/20) and by a grant from the United States-Israel Binational Science Foundation (BSF), Jerusalem, Israel, and the United States National Science Foundation.

\appendix
\section{Hetero-Nuclear Two-Channel Model}
\label{app:hetero}

The stand   ard two-channel model (one open atomic and one closed molecular channel) can be written for identical bosons (homo-nuclear) or distinguishable particles (hetero-nuclear).
While the latter is usually applied to a fermionic mixture of different spin states (Li-6 being a practical example~\cite{Huckans09}) it may of course be used for any distinguishable particles, such as different-spin bosons.

The hetero-nuclear two-body wave function, written here for $\sigma=3$, is
\begin{multline}
|\psi_{\text{het}}^{\left(\sigma=3\right)}\rangle=\beta_3\hat{d}_{3,\vec{q}=0}^{\dagger}|0\rangle
+\int\frac{d^{3}k}{\left(2\pi\right)^{3}}\alpha_3\left(\vec{k}\right)\hat{a}_{1,\vec{k}}^{\dagger}\hat{a}_{2,-\vec{k}}^{\dagger}|0\rangle.
\label{eq:2body:het:wave function}
\end{multline}
The Schr\"odinger equation $\hat{H}|\psi_{\text{het}}^{\left(\sigma=3\right)}\rangle=E|\psi_{\text{het}}^{\left(\sigma=3\right)}\rangle$, where $\hat{H}$ is given in Eq.~(\ref{eq:Hamiltonian:full}), leads to the following two coupled equations
\begin{subequations}
\begin{equation}
\beta_{3}\left(E_{\text{mol},3}-E\right)+\Lambda_{3}\int\frac{d^{3}k}{\left(2\pi\right)^{3}}\alpha_{3}\left(\vec{k}\right)=0
\end{equation}
\begin{equation}
\alpha_{3}\left(\vec{q}\right)\left(\frac{\hbar^{2}q^{2}}{m}+E_{12}-E\right)+\Lambda_{3}\beta_{3}=0
\end{equation}
\label{eq:2body:het:coupled equations}
\end{subequations}
The second equation shows that the free-particle amplitude $\alpha_{\sigma}\left(\vec{q}\right)$ is a Lorentzian of its argument $\vec{q}$, causing the integral in the first equation to diverge linearly in $\left|\vec{q}\right|$.
To deal with this UV divergence, as mentioned in the main text (Sec.~\ref{sec:2-body}), a high-momentum cut-off $k_{\text{cut-off}}$ is introduced.
The Lorentzian form of $\alpha_{\sigma}\left(\vec{q}\right)$ agrees with the expected exponential in position space (halo dimer).

In general, Eqs.~(\ref{eq:2body:het:coupled equations}) lead to three equivalent equations in the two-body sector due to the different coupling constants $\Lambda_\sigma$ and molecular and atomic detunings $E_{\text{mol},\sigma}$ and $E_\sigma$.
If one assumes, however, that $\Lambda_\sigma=\Lambda$, $E_{\text{mol},\sigma}=E_{\text{mol}}$ and $E_\sigma=0$ for all $\sigma$, the equations of each sector become identical.
The scattering amplitude $\tilde{f}\left(\tilde{k}_{0}\right)=-\tilde{\Lambda}\tilde{\beta}\left(\tilde{k}_{0}\right)/4\pi$ is then given by
\begin{equation}
\frac{1}{\tilde{f}\left(\tilde{k}_{0}\right)}=\frac{4\pi\left(\tilde{E}_{\text{mol}}-\tilde{k}_{0}^{2}\right)}{\tilde{\Lambda}^{2}}-\frac{2}{\pi}\left[1-\frac{i\pi}{2}\tilde{k}_{0}\right]
\end{equation}
and the dimer binding energy $-\tilde{\lambda}^2$ is found by solving
\begin{equation}
\left(\tilde{E}_{\text{mol}}+\tilde{\lambda}^{2}\right)-\frac{\tilde{\Lambda}^{2}}{2\pi^{2}}\left(1-\frac{\pi}{2}\tilde{\lambda}\right)=0.
\label{eq:2body:het:dimer}
\end{equation}
In the three-body sector one starts from
\begin{multline}
|\psi_{\text{het}}^{\left(\text{3B}\right)}\rangle=\sum_{\sigma}\int\frac{d^{3}q}{\left(2\pi\right)^{3}}\beta_{\sigma}\left(\vec{q}\right)\hat{d}_{\sigma,\vec{q}}^{\dagger}\hat{a}_{\sigma,-\vec{q}}^{\dagger}|0\rangle \\
+\int\frac{d^{3}q}{\left(2\pi\right)^{3}}\int\frac{d^{3}k}{\left(2\pi\right)^{3}}\alpha\left(\vec{q},\vec{k}\right)\hat{a}_{3,\vec{k}+\frac{\vec{q}}{2}}^{\dagger}\hat{a}_{1,-\vec{k}+\frac{\vec{q}}{2}}^{\dagger}\hat{a}_{2,-\vec{q}}^{\dagger}|0\rangle
\label{eq:3body:het:wave function}
\end{multline}
which, in general, leads to three coupled integral equations.
Assuming the three two-body subsystems to be identical reduces them to a single equation, namely
\begin{multline}
\left[\tilde{E}_{\text{mol}}+\frac{3\tilde{q}^{2}}{4}+\tilde{\lambda}^{2}-\frac{\tilde{\Lambda}^{2}}{2\pi^{2}}\left(1-\frac{\pi}{2}\sqrt{\frac{3\tilde{q}^{2}}{4}+\tilde{\lambda}^{2}}\right)\right]\tilde{\beta}\left(\tilde{q}\right) \\
-\frac{\tilde{\Lambda}^{2}}{2\pi^{2}}\int_{0}^{\infty}d\tilde{k}\ln\left(\frac{\tilde{k}^{2}+\tilde{k}\tilde{q}+\tilde{q}^{2}+\tilde{\lambda}^{2}}{\tilde{k}^{2}-\tilde{k}\tilde{q}+\tilde{q}^{2}+\tilde{\lambda}^{2}}\right)\tilde{\beta}\left(\tilde{k}\right)=0,
\label{eq:3body:het:equation for trimer energy}
\end{multline}
where the trimer binding energy is given by $-\tilde{\lambda}^2$.

For the populations in the open and closed channels of the two- and three-body sectors, see App.~\ref{app:populations}.

\section{Contrasting the Homo- and Hetero-Nuclear Two-Channel Models}
\label{app:homo and hetero}

The derivation in the homo and hetero-nuclear case are equivalent.
In particular, the identical two-body sectors assumption is trivial for homo-nuclear systems.
However, when acting with the interaction Hamiltonian $\sim\hat{d}^\dagger\hat{a}\hat{a}$ on the free particle wave function $\sim\hat{a}^\dagger\hat{a}^\dagger|0\rangle$, the homo-nuclear model allows for two paths for creating a molecule while in the hetero-nuclear model there is only one option.
This effect is known as Bose-enhancement.
The two- and three-body equations for the homo-nuclear model are thus obtained from the hetero-nuclear equations above by substituting $\tilde{\Lambda}\rightarrow\sqrt{2}\tilde{\Lambda}$.

Although Bose-enhancement is usually discussed in the context of many-body physics (e.g. the condensation of bosons into a BEC), its effect is already apparent at the two-body level and demonstrated beautifully by comparing the homo- and hetero-nuclear two-channel models.
From the above discussion we see that, if $\tilde{\Lambda}_{\text{het}}=\tilde{\Lambda}$, then $\tilde{\Lambda}_{\text{hom}}=\sqrt{2}\tilde{\Lambda}$, i.e. the effective coupling between the open channel and the closed channel is a factor $\sqrt{2}$ larger for identical bosons.
The stronger coupling leads to a broader Feshbach resonance, as compared to the hetero-nuclear model, with all of its characteristics: larger shift from the bare resonance, smaller $R^\star$, shallower dimer, slower population increase (decrease) of the closed (open) channel when moving away from resonance.

In the three-body sector, the larger effective coupling leads to a deeper bound trimer and merging features that are further away from resonance.
In fact, if one rescales both axes for the hetero-nuclear trimer in Fig.~\ref{fig:3body:trimer + population}(a) by a factor of $\left(\tilde{R}^\star_{\text{het}}/\tilde{R}^\star_{\text{hom}}\right)^2=4$, the two traces are identical.

\section{Degeneracy Limit of $\tilde{R}^\star_{bb}$ from $E_D$}
\label{app:limit of rStar}

In the main text we noted that, in the limit $E_{13}\rightarrow E_{22}$, the value of $\tilde{R}^\star_{bb}$ extracted from the dimer approaches $\tilde{R}^\star_{\text{hom}}\tilde{R}^\star_{\text{het}}/(\tilde{R}^\star_{\text{hom}}+\tilde{R}^\star_{\text{het}})$.

To derive this limit we start from Eq.~(\ref{eq:2body:dimer bb+ac}) for the dimer binding energy and plug in $E_{13}=E_{22}$.
Using $\tilde{R}^\star_{\text{hom}}=2\pi/\tilde{\Lambda_{22}^2}$ and $\tilde{R}^\star_{\text{het}}=4\pi/\tilde{\Lambda_{2}^2}$ one obtains
\begin{equation}
\left(\tilde{E}_{\text{mol},2}-\tilde{E}_{22}+\tilde{\lambda}^{2}\right)-\frac{2}{\pi}\left(\frac{1}{\tilde{R}^\star_{\text{hom}}}+\frac{1}{\tilde{R}^\star_{\text{het}}}\right)\left(1-\frac{\pi}{2}\tilde{\lambda}\right)=0.
\end{equation}
Comparison with Eq.~(\ref{eq:2body:het:dimer}) leads to the aforementioned expression for $\tilde{R}^\star_{bb}$.

\section{From the Two- to the Three-Body Sector}
\label{app:from 2 to 3 body}

In the main text we noted that $|\psi_{\text{3B}}\rangle$ in Eq.~(\ref{eq:3body:wave function}) has three different, uncoupled two-body subsystems, as illustrated in Fig.~\ref{fig:2body:connectivity}.
In fact, this effect is true also for the hetero-nuclear two-channel model whose three-body wave function, Eq.~(\ref{eq:3body:het:wave function}), has three two-body subsystems of the form of Eq.~(\ref{eq:2body:het:wave function}).
To illustrate how to go from the two- to the three-body wave function we introduce the operator
\begin{equation}
\hat{\mathcal{O}}_{\sigma=3}\left(\vec{q}\right)=\beta_{3}\left(\vec{q}\right)\hat{d}_{3,\vec{q}}^{\dagger}+\int\frac{d^{3}k}{\left(2\pi\right)^{3}}\alpha_{3}\left(\vec{q},\vec{k}\right)\hat{a}_{1,\vec{k}+\frac{\vec{q}}{2}}^{\dagger}\hat{a}_{2,-\vec{k}+\frac{\vec{q}}{2}}^{\dagger}
\end{equation}
with which Eq.~(\ref{eq:2body:het:wave function}) can be written as
\begin{equation}
|\psi_{\text{het}}^{\left(\sigma\right)}\rangle=\hat{\mathcal{O}}_{\sigma}\left(\vec{q}=0\right)|0\rangle.
\end{equation}
The operator $\hat{\mathcal{O}}_{\sigma}\left(\vec{q}\right)$ creates a two-body system of type $\sigma$ with center-of-mass momemtum $\hbar\vec{q}$.
We add a particle of type $\sigma$ and with momentum $-\hbar\vec{q}$ to $|\psi_{\text{het}}^{\left(\sigma\right)}\rangle$ and sum over $\sigma=1,2,3$ to obtain
\begin{equation}
|\psi_{\text{het}}^{\left(\text{3B}\right)}\rangle=\sum_{\sigma}\int\frac{d^{3}q}{\left(2\pi\right)^{3}}\hat{\mathcal{O}}_{\sigma}\left(\vec{q}\right)\hat{a}_{\sigma,-\vec{q}}^{\dagger}|0\rangle.
\end{equation}
Comparing this expression to Eq.~(\ref{eq:3body:het:wave function}) shows that the $\beta_\sigma$-amplitudes are the same in the two- and three-body sector, while the two-body $\alpha_\sigma$-amplitudes are related to the three-body $\alpha$-amplitude via
\begin{multline}
\alpha\left(\vec{q},\vec{k}\right)=\alpha_{1}\left(\vec{k}-\frac{\vec{q}}{2},-\frac{\vec{k}}{2}-\frac{3\vec{q}}{4}\right) \\
+\alpha_{2}\left(\vec{q},\vec{k}\right)+\alpha_{3}\left(-\vec{k}-\frac{\vec{q}}{2},-\frac{\vec{k}}{2}+\frac{3\vec{q}}{4}\right).
\end{multline}
In the homo-nuclear case the identity of the open channel atoms renders the $\sigma$ index redundant leading to the conclusion that the three-body $\alpha$-amplitude is equal to its two-body counter-part with finite center-of-mass momentum.


\section{Open and Closed Channel Populations in the Homo- and Hetero-Nuclear Model}
\label{app:populations}

We present expressions for the populations of the open and closed channels in the two- and three-body sector of the hetero-nuclear model using the equal two-body sector assumption.

In the two-body sector the population of the open and closed channels for the dimer are
\begin{subequations}
\begin{equation}
P_{\text{open}}=\frac{1}{\mathcal{N}}\int\frac{d^3k}{\left(2\pi\right)^3}\left|\alpha\left(\vec{k}\right)\right|^2
\end{equation}
\begin{equation}
P_{\text{closed}}=\frac{1}{\mathcal{N}}\left|\beta\right|^2,
\end{equation}
\end{subequations}
where
\begin{equation}
\mathcal{N}=\int\frac{d^3k}{\left(2\pi\right)^3}\left|\alpha\left(\vec{k}\right)\right|^2+\left|\beta\right|^2.
\end{equation}
Using the second equation in~(\ref{eq:2body:het:coupled equations}) one replaces $\alpha\left(\vec{k}\right)$ for $\beta$, which conveniently cancels, to obtain
\begin{subequations}
\begin{equation}
P_{\text{open}}=\frac{1}{\mathcal{N}}\left(\frac{\tilde{\Lambda}^{2}}{8\pi\tilde{\lambda}}\right)
\end{equation}
\begin{equation}
P_{\text{closed}}=\frac{1}{\mathcal{N}}
\end{equation}
\end{subequations}
and
\begin{equation}
\mathcal{N}=\frac{\tilde{\Lambda}^{2}}{8\pi\tilde{\lambda}}+1.
\end{equation}
In the three-body sector, similar to the two-body sector, the populations for the trimer are
\begin{subequations}
\begin{equation}
P_{\text{open}}=\frac{1}{\mathcal{N}}\int\frac{d^3q}{\left(2\pi\right)^3}\int\frac{d^3k}{\left(2\pi\right)^3}\left|\alpha\left(\vec{q},\vec{k}\right)\right|^2
\end{equation}
\begin{equation}
P_{\text{closed}}=\frac{1}{\mathcal{N}}\int\frac{d^3q}{\left(2\pi\right)^3}\left|\beta\left(\vec{q}\right)\right|^2,
\end{equation}
\end{subequations}
and $\mathcal{N}$ is the appropriate normalization factor (the sum of both integrals).
The free-particle amplitude $\alpha\left(\vec{q},\vec{k}\right)$ is expressed in terms of the molecule-atom amplitude $\beta\left(\vec{q}\right)$ via the Schr\"odinger equation upon which the integral over $\vec{k}$ can be solved.
The open channel population thus simplifies to
\begin{equation}
P_{\text{open}}=\frac{\tilde{\Lambda}^2}{8\pi\mathcal{N}}\int\frac{d^3\tilde{q}}{\left(2\pi\right)^3}\frac{\left|\beta\left(\tilde{q}\right)\right|^2}{\sqrt{3\tilde{q}^2/4+\tilde{\lambda}^2}}.
\end{equation}
Having found the trimer binding wave number and its wave function from Eq.~(\ref{eq:3body:het:equation for trimer energy}), the populations may be computed.

In contrast to the equations presented in App.~\ref{app:hetero}, here the homo-nuclear expressions are identical to those of the hetero-nuclear model.
However, when plotted as a function of $E_{\text{mol}}$ they differ as described at the end of App.~\ref{app:homo and hetero}, due to the difference in binding energy.

\section{Drivation of the Open and Closed Channel Populations in the Three-Body Sector}
\label{app:3body:populations}

According to the three-body wave function in Eq.~(\ref{eq:3body:wave function}), the open channel populations are ($i=$bbb, abc)
\begin{equation}
P_{i}=\frac{1}{\mathcal{N}}\int\frac{d^{3}q}{\left(2\pi\right)^{3}}\int\frac{d^{3}k}{\left(2\pi\right)^{3}}\left|\alpha_{i}\left(\vec{q},\vec{k}\right)\right|^{2}
\end{equation}
and the closed channel populations are ($\sigma=1,2,3$)
\begin{equation}
P_{\sigma}=\frac{1}{\mathcal{N}}\int\frac{d^{3}q}{\left(2\pi\right)^{3}}\left|\beta_{\sigma}\left(\vec{q}\right)\right|^{2}.
\end{equation}
The normalization constant $\mathcal{N}$ is found from $\sum_i P_i+\sum_\sigma P_\sigma=1$.
From the solution of Eqs.~(\ref{eq:3body:trimer coupled equations}) the binding wave number $\lambda$ and the eigen vectors $\beta_\sigma$ are obtained.
The closed channel populations are thus readily computed.
For the open channel populations we use the Schr\"odinger equations to express $\alpha_i$ in terms of $\beta_\sigma$, resulting in
\begin{equation}
P_{\text{bbb}}=\frac{1}{\mathcal{N}}\int\frac{d^{3}\tilde{q}}{\left(2\pi\right)^{3}}\int\frac{d^{3}\tilde{k}}{\left(2\pi\right)^{3}}\left|\frac{\tilde{\Lambda}_{22}\beta_{2}\left(\tilde{q}\right)}{\tilde{k}^{2}+3\tilde{q}^{2}/4+\tilde{\lambda}^{2}}\right|^{2}
\end{equation}
for the $bbb$-open-channel.
Since $\beta_{2}\left(\tilde{q}\right)$ is $k$-independent the integral over the latter may be computed and we get
\begin{equation}
P_{\text{bbb}}=\frac{\tilde{\Lambda}_{22}^{2}}{8\pi\mathcal{N}}\int\frac{d^{3}\tilde{q}}{\left(2\pi\right)^{3}}\frac{\left|\beta_{2}\left(\tilde{q}\right)\right|^{2}}{\sqrt{3\tilde{q}^{2}/4+\tilde{\lambda}^{2}}}.
\end{equation}
For $i=$abc the substitution leads to
\begin{multline}
P_{\text{abc}}=\frac{1}{\mathcal{N}}\int\frac{d^{3}\tilde{q}}{\left(2\pi\right)^{3}}\int\frac{d^{3}\tilde{k}}{\left(2\pi\right)^{3}} \\
\times\left|\frac{\tilde{\Lambda}_{1}\beta_{1}\left(\vec{k}-\frac{\vec{q}}{2}\right)+\tilde{\Lambda}_{2}\beta_{2}\left(\tilde{q}\right)+\tilde{\Lambda}_{3}\beta_{3}\left(\vec{k}+\frac{\vec{q}}{2}\right)}{\tilde{k}^{2}+3\tilde{q}^{2}/4+\tilde{E}_{123}-\tilde{E}_{222}+\tilde{\lambda}^{2}}\right|^{2}.
\end{multline}
At this stage we make the simplifying assumption $\beta_{1,3}\ll\beta_{2}$, which is equivalent to $P_{1,3}\ll P_2$, permitting integration over $k$:
\begin{equation}
P_{\text{abc}}=\frac{\tilde{\Lambda}_{2}^{2}}{8\pi\mathcal{N}}\int\frac{d^{3}\tilde{q}}{\left(2\pi\right)^{3}}\frac{\left|\beta_{2}\left(\tilde{q}\right)\right|^{2}}{\sqrt{3\tilde{q}^{2}/4+\tilde{E}_{123}-\tilde{E}_{222}+\tilde{\lambda}^{2}}}.
\end{equation}
While the expressions for $P_{\text{bbb}}$ and $P_\sigma$ are exact, the expression for $P_{\text{abc}}$ is approximate.
The approximation breaks down when $P_{1,3}\approx P_2$ which is the case at $E_{\text{mol}}>E_{\text{res}}$ [see Fig.~\ref{fig:3body:trimer + population}(b,c)].
The populations at $a_-$ are thus not captured well by the expressions derived here.


\begin{thebibliography}{10}
\bibitem{bb_channel_meaning}
The `$bb$-channel' refers to two free atoms in the $b$-state. By `scattering in
  the $bb$-channel' we mean a scattering event for which both the initial and
  the final state consist of two separated atoms in the $b$-state. However, the
  intermediate state is a superposition of all $M_{\text{tot}}=0$ two-body
  channels. A loosely bound dimer associated with a Feshbach resonance in the
  $bb$-channel has $M_{\text{tot}}=0$ spin character and its wavefunction has
  projections on multiple channels. At the resonance position the projection on
  the $bb$-channel approaches unity and the dimer is thus called the
  `$bb$-channel dimer'. Note the somewhat confusing jargon: Our model of the
  $bb$-channel, which describes scattering and dimers in the sense explained
  above, includes the $bb$-channel (two atoms in the $b$-state), the
  $ac$-channel and a closed channel made up of the remaining $M_{\text{tot}}=0$
  two-body channels.
\bibitem{ccOnlySWave}
Throughout this paper, coupled-channels calculations were performed using an
  $s$-wave basis only.
\bibitem{Li7_rescaleAxis}
It is possible to produce very similar $aa$ and $bb$ scattering lengths (except
  for the region of the narrow resonance) by rescaling the magnetic field axis
  of the $aa$-channel by a factor of $1.18$. In this case, also the two dimers
  lie on top of each other, suggesting that the two resonances are identical.
  However, this procedure and conclusion is problematic for at least two
  reasons. First, due to repulsion from the narrow resonance dimer, if both
  resonances are identical, the $bb$ dimer should be deeper. The fact that it
  is not, even after rescaling, points to the shallowing effect of the
  $ac$-channel. Second, rescaling the magnetic field axis physically means
  changing the magnetic moment of the bare molecule. However, since the two
  resonances arise due to the same bare molecule of the singlet potential, such
  a change is unphysical.
\bibitem{sinEtaIskf}
The result of the coupled-channels calculations is the scattering matrix
  $S_{ij}$, which describes a scattering event from channel $i$ to channel $j$,
  e.g. $bb\rightarrow bb$ or $bb\rightarrow ac$. From $S_{ij}$ one may compute
  $T_{ij}=\delta_{ij}-S_{ij}$ and $\sin^2\eta_{ii}=\left|T_{ii}\right|^2/4$. In
  the case of $s$-wave scattering, the $i\rightarrow i$ scattering amplitude is
  $f_{ii}=-T_{ii}/2ik$. Thus, $\sin^2\eta_{ii}=\left|kf_{ii}\right|^2$ and
  Fig.~\ref{fig:Li7:scattering amplitude} is directly comparable to
  Figs.~\ref{fig:2body:scattering amplitude on linear scale}
  and~\ref{fig:2body:scattering amplitude}.
\bibitem{3body_aad_channel}
As mentioned in the beginning of Sec.~\ref{sec:model + hamiltonian}, here we
  deal with a $F=1$ spinor gas. However, the ground state of $^7$Li has both a
  $F=1$ and $F=2$ part (Fig.~\ref{fig:Li7:Zeeman shift}). The energetically
  low-lying three-body continua consist of three particles with $J=-1/2$. There
  are three options to form such a continuum, namely, $bbb$, $abc$
  ($E_{abc}-E_{bbb}=28$~MHz) and $aad$ ($E_{aad}-E_{bbb}=108$~MHz). The latter
  is neglected because, first, it does not exist in a $F=1$ spinor gas and,
  second, because, albeit close, it is almost four times further detuned than
  the leading order $abc$ contribution.
\end{thebibliography}
\bibliographystyle{unsrt}

\end{document}